\documentclass[fleqn,usenatbib,useAMS]{aa}  
\usepackage{xspace}

\usepackage{lmodern}
\usepackage{amsmath}	
\usepackage{amssymb}
\usepackage{hyperref}
\usepackage{graphicx}
\usepackage{inputenc}
\usepackage[T1]{fontenc}
\usepackage{epsfig}
\usepackage{natbib}
\usepackage{relsize}
\usepackage{ae,aecompl}
\usepackage{graphicx}	
\usepackage{times}
\usepackage{subcaption}
\def\kms{km\,s$^{-1}$}
\def\lya{Ly$\alpha$}

\def\fesc{$f_\mathrm{esc}^{LyC}$~}
\def\fescp{$f_\mathrm{esc}^{LyC}$}
\def\fescly{$f_\mathrm{esc}^{Ly\alpha}$~}
\def\fesclyp{$f_\mathrm{esc}^{Ly\alpha}$}
\def\hi{\ion{H}{i}}

\def\ha{H$\alpha$}

\newcommand{\vcenp}{$v_{\text{cen}}$}
\newcommand{\vnp}{$v_{90}$}
\newcommand{\vcen}{$v_{\text{cen}}$~}
\newcommand{\vn}{$v_{90}$~}
\newcommand{\mstar}{$M_\ast$ }
\newcommand{\mstarp}{$M_\ast$}
\newcommand{\sfrsd}{$\Sigma_{\text{SFR}}$ }
\newcommand{\sfrsdp}{$\Sigma_{\text{SFR}}$}
\newcommand{\kmsp}{km~s$^{ -1}$}
\newcommand{\oi}{\ion{O}{i}~}

\newcommand{\siii}{\ion{Si}{ii}~}
\newcommand{\siiii}{\ion{Si}{iii}~}

\newcommand{\siiv}{\ion{Si}{iv}~}

\newcommand{\siiip}{\ion{Si}{ii}}

\newcommand{\siiiip}{\ion{Si}{iii}}

\newcommand{\oiiip}{\ion{O}{iii}}

\newcommand{\oiip}{\ion{O}{ii}}

\newcommand{\siivp}{\ion{Si}{iv}}

\newcommand{\sfr}{M$_\odot$~yr$^{-1}$ }
\newcommand{\sfrp}{M$_\odot$~yr$^{-1}$}

\newcommand{\msun}{M$_\odot$}

\begin{document}

   \title{Do galaxies that leak ionizing photons have extreme outflows?
   }

 \author{John Chisholm\inst{1},
I. Orlitov\'a\inst{2},
D. Schaerer\inst{1,3}, 
A. Verhamme\inst{1},
G. Worseck\inst{4}, 
Y. I. Izotov$^{5}$,
T. X. Thuan$^{6}$, and N.~G.~Guseva$^{5}$
  }
 \offprints{John.Chisholm@unige.ch}
  \institute{
Observatoire de Gen\`eve, Universit\'e de Gen\`eve, 51 Ch. des Maillettes, 1290 Versoix, Switzerland
         \and
Astronomical Institute, Czech Academy of Sciences, Bo\v cn{\'\i} II 1401, 141 00, Prague, Czech Republic
         \and
CNRS, IRAP, 14 Avenue E. Belin, 31400 Toulouse, France
        \and
Max-Planck-Institut f\"ur Astronomie, K\"onigstuhl 17, 69117 Heidelberg, Germany             
        \and
Main Astronomical Observatory, Ukrainian National Academy of Sciences,
27 Zabolotnoho str., Kyiv 03143, Ukraine
        \and
Astronomy Department, University of Virginia, P.O. Box 400325, 
Charlottesville, VA 22904-4325, USA
         }

\authorrunning{J. Chisholm et al.}
\titlerunning{Outflows in LyC leaking galaxies}


 
  \abstract
   {To reionize the early universe, high-energy photons must escape the galaxies that produce them. How these photons escape is debated because too many ionizing photons are absorbed even at small \ion{H}{i} column densities. It has been suggested that stellar feedback drives galactic outflows out of star-forming regions, creating low density channels through which ionizing photons escape into the inter-galactic medium.}{We compare the galactic outflow properties of confirmed Lyman continuum (LyC) leaking galaxies to a control sample of nearby star-forming galaxies to explore whether the outflows from leakers are extreme as compared to the control sample.}{We use data from the Cosmic Origins Spectrograph on the {\sl Hubble Space Telescope} to measure the equivalent widths and velocities of \siii and \siiii absorption lines, tracing neutral and ionized galactic outflows. We explore whether the leakers have similar outflow properties to the control sample, and whether the outflows from the leakers follow similar scaling relations with host galaxy properties as the control sample. We rederive the escape fraction of ionizing photons for each leaker, and study whether the outflow properties influence  the LyC escape fractions.}{We find that the \siii and \siiii equivalent widths of the LyC leakers reside on the low-end of the trend established by the control sample. The leakers' velocities are not statistically different than the control sample, but their absorption line profiles have a different asymmetry: their central velocities are closer to their maximum velocities. This possibly indicates a more rapidly accelerated outflow due to the compact size of the leakers. The outflow kinematics and equivalent widths are consistent with the scaling relations between outflow properties and host galaxy properties -- most notably metallicity -- defined by the control sample. Additionally, we use the Ly$\alpha$ profiles to show that the \siii equivalent width scales with the Ly$\alpha$ peak velocity separation.}{We determine that the low equivalent widths of the leakers are likely driven by low metallicities and low \ion{H}{i} column densities, consistent with a density-bounded ionization region, although we cannot rule out significant variations in covering fraction. While we do not find that the LyC leakers have extreme outflow velocities, the low maximum-to-central velocity ratios demonstrate the importance of the acceleration and density profiles for LyC and Ly$\alpha$ escape.}


   \maketitle
%
\section{Introduction}

Neutral hydrogen absorbs and scatters photons at 1216\AA\ (\lya). Therefore, intergalactic \hi\ absorbs photons in the Lyman series along the line-of-sight to distant galaxies. However, we observe distant galaxies at these wavelengths, implying that the inter-galactic hydrogen must be mostly ionized. Since hydrogen recombines early in the history of the universe to form the cosmic microwave background radiation, an intermediary period exists where hydrogen is reionized. Recent polarization observations of the cosmic microwave background radiation suggest that reionization finishes near redshifts of 6-8 \citep{planck, greig16}.  After reionization, intergalactic hydrogen is highly ionized \citep[neutral fraction $\sim$10$^{-4}$; ][]{fan2006}, with the remaining neutral gas in damped Ly$\alpha$ systems \citep{crighton15, sanchez-ramirez16}. 

Exactly how the universe became reionized is hotly debated. Quasars are an attractive way to emit high energy photons because gas funneling onto a super-massive black hole is heated to high temperatures, which radiates copious amounts of ionizing photons. Consequently, simulations suggest that quasars could dominate the reionization budget \citep{Madau15}. However, the quasar density function falls off rapidly with redshift \citep{hopkins08}, such that at high redshifts it is unlikely that there are enough quasars to reionize the universe \citep{Fontanot12}, implying that AGN contribute at most 30\% of the required ionizing photons \citep{ricci}. Consequently, studies focus on star-forming galaxies, but to reionize the universe up to 20\% of the total ionizing photons produced by massive stars must leak out of galaxies \citep{Ouchi09, Robertson13, robertson15, Dressler15}. Observations rarely find this high of escape fractions \citep{Grazian16, Rutkowski16}. 

\begin{table*}
\centering
\caption{Host galaxy properties for the  nine LyC leakers.}
\begin{tabular}{lcccccccc}
\hline
\hline
Galaxy Name & log(\mstarp) & SFR & log(O/H) + 12 & $z$ &  O$_{32}$ & $R$ & \fesc & References \\
 & (log(M$_\odot$)) & (M$_\odot$ yr$^{-1}$) & (dex) & & & (kpc)& (\%) &  \\
\hline 
(1) & (2)  & (3) & (4) & (5) & (6) & (7) & (8) & (9)  \\
\hline 
J1152$+$3400 & 9.4$^6$  & 24$^6$ & 8.0$^6$ & 0.3419 &  5.4$^6$ & 0.37 & 13 & \citet{Izotov16b}\\
J0925$+$1403 & 8.7$^5$ & 32$^5$ & 7.9$^5$& 0.3013 & 4.8$^5$ & 0.28 & 7 & \citet{Izotov16, Izotov16b}\\
J1442$-$0209 & 8.8$^6$ & 22$^6$ & 7.9$^6$ & 0.2937 &  6.7$^6$ & 0.25 & 7& \citet{Izotov16b}\\
J1333$+$6246 & 8.3$^6$ & 9$^6$ & 7.8$^6$  & 0.3181 &   4.8$^6$ & 0.40 & 6& \citet{Izotov16b}\\
J1503$+$3644 & 8.0$^6$ & 24$^6$ & 8.0$^6$ & 0.3557 & 4.9$^6$ & 0.38 & 6& \citet{ Izotov16b}\\
Haro~11 & 10.1$^3$ & 26$^3$ & 8.1$^7$ &0.0206  & 1.5$^{12}$ & 2.19 & 3 & \citet{Alexandroff15}\\
Tol~0440$-$381 & 10.0$^4$ & 3.0$^{10}$  & 8.2$^2$ & 0.0410 & 2.0$^{12}$ & & <2 & \citet{Leitherer16}\\
J0921$+$4509 & 9.9$^3$ & 77$^3$ & 8.7$^1$ & 0.2350 & 0.3$^{12}$  &0.78 & 1 & \citet{Borthakur14} \\
Tol~1247$-$232 & 9.7$^9$ & 16$^{10}$  & 8.1$^{11}$ & 0.0482&  3.4$^{12}$  & & <0.4 & \citet{Leitherer16}\\
\hline
\end{tabular}
\tablefoot{Column (1) is the name of the galaxy, column (2) is the stellar mass (\mstarp), column (3) is the star formation rate (SFR), column (4) is oxygen-to-hydrogen abundance (log(O/H)+12), column (5) is the redshift ($z$), column (6) is the O$_{32}$ = [O {\sc iii}] 5007\AA/[O {\sc ii}]~3727\AA\ flux ratio, column (7) is 50\% of the Petrosian radius measured from the COS acquisition images ($R$), and column (8) is the LyC escape fraction (\fescp). $R$ is not measured for the two \citet{Leitherer16} galaxies because COS images are not available. We include the FUSE galaxy, Haro~11 as a LyC candidate. Notes on the re-measurement of the \fesc from the COS spectra are given in the Appendix. Column (9) gives previous references for the COS spectra of each galaxy. The references for the individual parameters are: $^1$\citet{Borthakur14}, $^2$\citet{campbel1986}, $^3$\citet{Chisholm15}, $^4$\citet{Heckman11}, $^5$\citet{Izotov16}, $^6$\citet{Izotov16b}, $^7$\citet{James14} $^8$\citet{kinney1993}, $^9$\citet{Leitet13}, $^{10}$\citet{Leitherer16}, $^{11}$\citet{terlevich1993}, $^{12}$\citet{verhamme16}.}
\label{tab:sample}
\end{table*}

A possible reconciliation is that we do not know how many galaxies leak Lyman continuum (photons with wavelengths less than 912\AA; LyC) photons. Recent studies emphasize low-mass galaxies as the source of reionization because the faint-end of the ultraviolet luminosity function rises steeply in the early universe \citep{yan04, Finkelstein11, oesch14, bouwens15}. A steeper luminosity function provides more sources of ionizing photons, and reduces the fraction needed to escape. However, it is observationally unconstrained how many ionizing photons escape galaxies, and how the escape fraction varies with galaxy properties. 

Measuring the LyC escape fraction (\fescp) is quite challenging because it requires deep restframe far-ultraviolet observations. There are only a few confirmed high redshift Lyman continuum leakers with relative \fesc between 20-65\%  \citep{Vanzella15, deBarros16, Vanzella16, shapley16, bian}, while most investigations only place upper limits on \fesc between 5-50\% \citep{Vanzella10, bouwens15, sandberg15, Siana15}.

A particular hurdle of high redshift LyC observations is contamination from foreground galaxies \citep{Vanzella10}. Additionally, the \lya\ forest absorbs most of the ionizing photons above redshifts of 4, making it impossible to study the LyC of high-redshift galaxies. The Cosmic Origins Spectrograph (COS) on the {\sl Hubble Space Telescope} \citep[\sl HST;][]{cos} has increased sensitivity and precision over previous ultraviolet spectrographs, enabling studies to focus on nearby galaxies as reionization analogs. However, only 9 local LyC leakers have been detected so far \citep{Bergvall06, Leitet13, Borthakur14, Izotov16, Izotov16b, Leitherer16}. Why do so few local galaxies emit LyC photons?

Extreme galactic outflows, characterized by larger than average velocities and mass outflow rates, in the early universe might allow ionizing photons to escape \citep{Heckman11}. Energy and momentum from star formation launches gas out of star-forming regions as large-scale galactic outflows \citep{heckman2000, veilleux2005, erb14}. If early galaxies have extreme galactic outflows, then outflows remove neutral gas from star-forming regions, and create holes through which ionizing photons pass \citep{Heckman11, Jones13, Alexandroff15, sharma}. This \lq{}\lq{}picket-fence\rq{}\rq{} model manifests itself with low covering fractions of neutral absorption lines, like \oi or \siii \citep{Alexandroff15}. A second way for LyC photons to escape is for young star clusters to ionize the surrounding \hi, and produce regions where the \hi\ density truncates before the ionizing photons are completely absorbed \citep{jaskot13, Nakajima14}. Leakage though these \lq{}\lq{}density-bounded\rq{}\rq{} regions manifests itself through high ionization ratios ([\oiiip]/[\oiip]) and low \hi\ column densities \citep{jaskot13, Nakajima14}.

Do galaxies that leak ionizing photons have extreme outflows? This question drives this paper, and the answer will help determine how LyC photons leak from galaxies. To answer it, we compile a sample of local LyC leakers and compare their neutral and ionized outflow properties, as measured from the \siii and \siiii absorption lines, to a sample of local star-forming galaxies \citep{Chisholm16}. This comparison allows us to determine whether LyC leakers have different average outflow properties than galaxies with unknown, although likely low, \fescp. Specifically, we test whether leakers have velocities and equivalent widths, a proxy for the strength of the outflow, that are larger than the average outflow from galaxies with similar star formation rates and stellar masses. This comparison tests whether the leakers exhibit extremely high-velocity or strong outflows. In Sect.~\ref{data} we describe the leaker and control samples; and how we measure the host galaxy, outflow, and  \lya\ properties. We compare the outflow properties of the leakers to the outflow properties, host galaxy properties, and \lya\ properties of the control sample (Sect.~\ref{results}). Then we compare our results to previous studies of low redshift (Sect.~\ref{comp}) and high redshift galaxies (Sect.~\ref{highz}). We finish by discussing how the outflow strengths (Sect.~\ref{ew}), the velocity distributions, and the density distributions impact LyC escape (Sect.~\ref{rat}). Our main conclusions are summarized in Sect.~\ref{summary}.

\section{Data}
\label{data}
\subsection{Sample Selection}
To compare the outflow properties of LyC leaking galaxies to normal star-forming galaxies, we compose three samples: a sample of known LyC emitting galaxies (hereafter called the leaker sample), a sample of local star-forming galaxies with largely unknown LyC properties (hereafter called the control sample), and a sample of \lya\ emitting galaxies (hereafter called the \lya\ sample).  The \lya\ sample is created by combining galaxies with measured \lya\ properties from (1) the Lyman Alpha Reference Sample \citep[LARS;][]{Hayes13, Ostlin14} that overlap with the control sample and (2) the Green Pea sample from \citet{Henry15}. This \lya\ sample is used in later sections to expand the dynamic range of the \hi\ column density, demonstrating important correlations between outflow properties and \hi\ column densities. 

The control sample is composed of the 26 local, star-forming galaxies from \citet{Chisholm16} that have restframe UV spectra from COS on {\sl HST}, along with one galaxy from \citet{Leitherer16} that does not have a LyC detection. Each control sample galaxy has a \siiip~1260\AA\ and \siiiip~1206\AA\ outflow, which is defined as having a central velocity less than 0~\kms\ at the 1$\sigma$ significance level. The control sample has a median stellar age of 5.6~Myr, star formation rates between 0.01 and 136~M$_\odot$~yr$^{-1}$, stellar masses between 10$^7$~M$_\odot$ and 10$^{11}$~M$_\odot$, and samples galaxies of all sizes, from small compact Green Peas to large extended spiral galaxies \citep{Chisholm15}.

While the LyC properties of the control sample are unknown, previous studies suggest that the escape fraction is likely low in these galaxies \citep{Rutkowski16, Grazian16}. We can quantify whether the control sample is heavily contaminated by LyC leakers in two ways. First, one galaxy from the \citet{Leitherer16} sample, Mrk~54, has a \fesc consistent with zero (see the  Appendix). We include this galaxy in the control sample, and it has outflow properties that are consistent with the control sample (\siii equivalent width of 1.3\AA\ and \vn of $-655$~\kms; see below). Moreover, six of the control sample galaxies (NGC~3690, NGC~4214, NGC~5253, NGC7714, M~83, and 1~Zw~18) were observed with the {\it Far Ultraviolet Space Telescope} (FUSE), and none of these galaxies have measurable escape fractions. \citet{Heckman11} place a median upper limit for their individual escape fractions at 1\%, indicating that the control sample likely has low levels of LyC escape. Even though \fesc is unknown for three-quarters of the control sample, comparison with the leaker sample remains useful because it describes whether leakers have faster and more massive outflows than average local star-forming galaxies.

We acquired the restframe UV spectra of the leaker sample from four recent COS studies that measured the LyC escape fraction: five from {\sl HST} proposal ID 13744 \citep[][]{Izotov16, Izotov16b}, two from {\sl HST} proposal ID 13325 \citep[][]{Leitherer16}, one from {\sl HST} proposal ID 12886  \citep[][]{Borthakur14}, and Haro~11 from {\sl HST} proposal ID 13017  \citep{Alexandroff15, Heckman15}. However, the LyC detection for Haro~11 comes from FUSE observations \citep{Leitet13}, and the \fesc has not been confirmed with COS. In the Appendix, we caution that the \citet{Leitherer16} points (red diamonds in all plots) could be contaminated by geocoronal Lyman series emission, and the \fesc may be over-estimated. Whether these points are included does not change the overall conclusions drawn about the outflow properties from leakers.

These are the nine known local LyC leaking galaxies from the literature, spanning redshifts between 0.02 and 0.36. These galaxies tend to have large [\ion{O}{iii}]~5007\AA/[\ion{O}{ii}]~3727\AA\ ratios (hereafter O$_{32}$) and compact sizes, as described in \citet{Izotov16b}. We follow \citet{Alexandroff15} by measuring the galaxy sizes as 50\% of the Petrosian radii, as determined from the COS acquisition images. The LyC leakers tend to have compact star forming regions, with sizes between $0.3-2.2$~kpc (Table~\ref{tab:sample}).

\subsection{Host galaxy properties}
The host galaxy properties of the leakers are drawn from the literature, and listed in Table~\ref{tab:sample}. The stellar masses (\mstarp), metallicities, and star formation rates (SFR) for the control sample are drawn from \citet{Chisholm16}, where \mstar is calculated using a Chabrier initial mass function \citep[IMF;][]{chabrier} and archival Wide-field Infrared Survey Explorer ({\sl WISE}) observations \citep{Jarrett2013, querejeta}. The SFRs of the control sample are calculated using a combination of the FUV fluxes from the Galaxy Evolution Explorer ({\sl GALEX}) and 22~$\mu$m fluxes from {\sl WISE}, using the calibration from \citet{Buat11}. This SFR method includes both IR and UV star formation, while accounting for dust heating by evolved stars. Since different methods are used to calculate the SFR, slight biases may arise if the star formation is extremely bursty \citep{izotov14}. For the leakers, \mstar and SFR are tabulated in Table~\ref{tab:sample}, after converting to a Chabrier IMF. Additionally, we compile the log(O/H)+12 values for the leaker sample from different literature sources, as noted in Table~\ref{tab:sample}.

\begin{figure}[h]
\includegraphics[width = 0.5\textwidth]{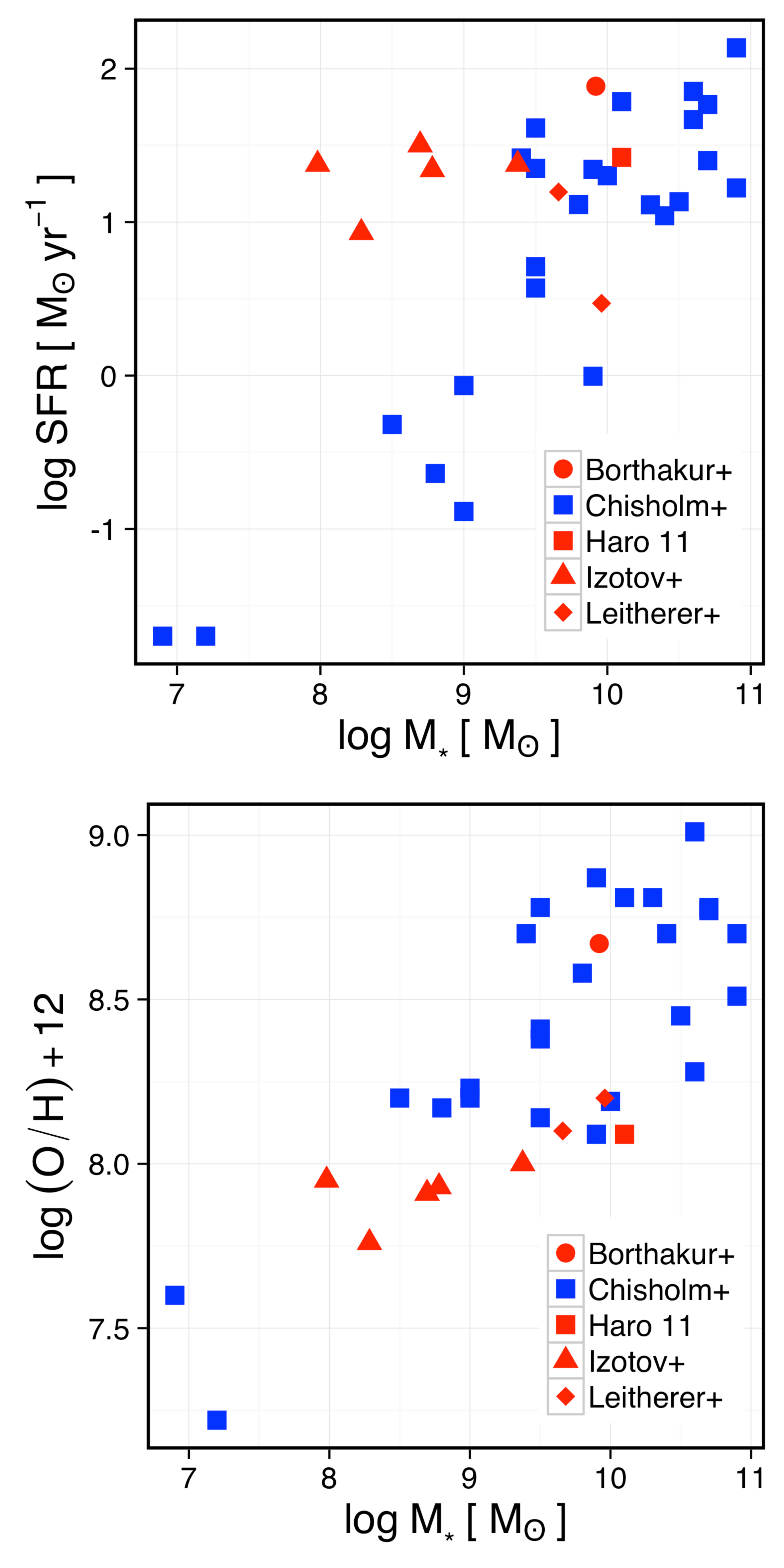}
\caption{{\it Upper panel:} Comparison of the star formation rate (SFR) with the stellar mass of the galaxy (\mstarp) for the leaker sample (red symbols) and the control sample (blue squares). The leakers generally have higher SFRs, at a given \mstarp, than the control sample. {\it Bottom panel:} The metallicity (log(O/H)+12) versus \mstarp. The leakers generally have lower metallicities than the control sample. Note that we only measure LyC upper-limits for the \citet{Leitherer16} sample (red diamonds; see the Appendix). }
\label{fig:mainseq}
\end{figure}

The \mstar and SFR values are shown in the upper panel of Fig.~\ref{fig:mainseq}, with the leaker sample shown as red symbols and the control sample shown as blue squares. The leakers have a narrow SFR range between 10 and 100~\sfrp, along with a narrow \mstar distribution between 10$^8$ and $10^{10.5}$ \msun. Many leakers have SFRs an order of magnitude larger than galaxies with similar \mstar in the control sample. These elevated SFRs place many leakers substantially off the main-sequence, with larger specific star formation rates ($sSFR=SFR/M_\ast$). Additionally, the lower panel of Fig.~\ref{fig:mainseq} shows the mass-metallicity relation for our galaxies. The leakers generally have lower metallicity than control sample galaxies at similar \mstarp. These offsets from the control sample are largely a selection effect because the \citet{Izotov16b} galaxies were selected to have large $O_{32}$, biasing the galaxies towards higher sSFRs and lower metallicities \citep{sanders}. All three of the host galaxy properties (\mstarp, SFR, and metallicity) are highly correlated (4-5$\sigma$), as expected from the fundamental metallicity relation \citep{mannucci2010}. 

\subsection{Data reduction}
Here we discuss the data reductions for the different samples. Some samples are observed with different instrument configurations, so we discuss the reduction of the leakers first (Sect.~\ref{lyc_red}) and then the control sample (Sect.~\ref{pre_red}). 

\subsubsection{Leakers}
\label{lyc_red}
The COS leakers' spectra were reduced with CALCOS~v2.21 and custom software for faint COS targets \citep{Worseck16}, and the data reduction is largely described in \citet{Izotov16, Izotov16b}. The leakers not in \citet{Izotov16b} were reduced analogously, except that (i) we employed custom pulse height filtering on both COS detector segments, and (ii) we adjusted the COS extraction aperture to include all continuum flux of the more extended galaxies \citep{Leitherer16}, and their extended \lya\ emission. The pulse height filtering maximizes the sensitivity and removes low-pulse-height blemishes on the COS detector segment B containing the LyC of the galaxies. Since the pulse height distribution changes with time \citep[e.g. ][]{Worseck16}, the pulse height range was determined for each spectra individually. For the G160M spectrum of J0921+4509 (Program 11727, PI: Heckman) we included pulse heights 2--15 on both segments, whereas for the G140L spectrum (Program 12886, PI: Borthakur) we chose pulse heights 2--15 on segment A and 2--14 on segment B. All data from Program 13325 (PI: Leitherer) were reduced with pulse height thresholds 1--15 on segment A and 3--16 on segment B, respectively.

Spectra were extracted in a rectangular aperture with a width that accounts for the astigmatism of the COS optics and the apparent extent of the target \citep{cos}. In particular, for the compact galaxy J0921+4509 we used a 25-pixel aperture that includes all flux from a point source, while the more extended galaxies from \citet{Leitherer16} were extracted with a 43-pixel aperture. We caution that absolute fluxing of extended targets is non-trivial due to the vignetting of the COS aperture. The detector dark current in the chosen aperture was estimated from dark calibration exposures taken within $\pm$1.5 months from the observations in similar orbital conditions, detector voltages, and pulse height cuts. Similar to \citet{Worseck16}, we verified the dark current estimation procedure on the dark exposures (Appendix below). The very small quasi-diffuse Galactic and extragalactic background was subtracted using the {\sl GALEX} far-ultraviolet map from \citet{murthy}. The impact of geocoronal Lyman series emission and scattered light was estimated by comparing the fluxes obtained in the total exposure time and in the orbital shadow.

The leakers from \citet{Izotov16, Izotov16b} were observed with two COS gratings, the G160M and the G140L, but the signal-to-noise ratio of the G160M grating is too low to analyze the silicon absorption features. To further increase the signal-to-noise ratio, we degrade the G160M data and combine it with the G140L observations at wavelengths where the data overlap (observed wavelengths of $1400-1800$~\AA) using the {\small IRAF} task {\it splice}. We achieve a final signal-to-noise ratio in the continuum of $\sim$2. The \citet{Leitherer16} galaxies are observed with the G140L grating and have similar spectral resolution as the \citet{Izotov16, Izotov16b} spectra. Meanwhile, Haro~11 and J0921$+$4509 are observed with the G130M grating, and we degrade the observations by convolving with a Gaussian that has a velocity width of 75~\kmsp, approximately the resolution of the G140L observations. This deconvolution is only approximate because the COS line spread function is highly non-Gaussian \citep{cos}. 

The combined observations have lower spectral resolution (from $R\sim 20,000$ to $R\sim 2,000$). With this decreased resolution, we cannot analyze narrow features or quantities that are likely resolved out, like column density or covering fraction, but the higher signal-to-noise ratio allows us to measure the velocities and equivalent widths of the silicon absorption lines.

\subsubsection{Control Sample}
\label{pre_red}
\begin{table}
\caption{The \siiip~1260\AA\ outflow properties for the nine confirmed LyC leakers.}
\begin{tabular}{lccc}
\hline
\hline
Galaxy Name & \siii EW & \siii \vcen & \siii \vn\\
 & (\AA) & (\kmsp) & (\kmsp) \\
\hline 
\multicolumn{1}{c}{(1)} & (2) & (3) & (4) \\
\hline 
J1152$+$3400 & $0.49 \pm 0.10$ & $-336 \pm 129$ & $-563 \pm 33$ \\
J0925$+$1403 & $0.75 \pm 0.15$ & $-64 \pm 48$ & $-244 \pm 14$ \\
J1442$-$0209 & $0.65 \pm 0.13$ & $-473 \pm 77$ & $-712 \pm 34$ \\
J1333$+$6246 & $0.75 \pm 0.16$ & $-237 \pm 145$ & $-461 \pm 41$ \\
J1503$+$3644 & $0.56 \pm 0.11$ & $-187 \pm 79$ & $-324 \pm 23$ \\
Haro~11 & $1.04 \pm 0.21$ & $-201 \pm 8$ & $-514 \pm 20$ \\
Tol~0440$-$381 & $0.21 \pm 0.02$ & $-66 \pm 14$ & $-100 \pm 35 $ \\
J0921$+$4509 & $0.67 \pm 0.07$ & $-50 \pm 8$ & $-382 \pm 10$ \\
Tol~1247$-$232 & $0.59 \pm 0.05$ & $-62 \pm 8$ & $-433 \pm 8$ \\
\hline
\end{tabular}
\tablefoot{Column (1) gives the name of the galaxy, column (2) gives the \siiip~1260\AA\ restframe equivalent width (EW), column (3) gives the velocity at 50\% of the total equivalent width (\vcenp), and column (4) gives the velocity at 90\% of the continuum (\vnp). The galaxies are ordered in terms of descending \fesc (see Table~\ref{tab:sample}).}
\label{tab:si2}
\end{table}

\begin{table}
\caption{The \siiiip~1206\AA\ outflow properties for the 7 LyC leakers with \siiii coverage.}
\begin{tabular}{lccc}
\hline
\hline
Galaxy Name & \siiii EW & \siiii \vcen & \siiii \vn\\
 & (\AA) & (\kmsp) & (\kmsp) \\
\hline 
\multicolumn{1}{c}{(1)} & (2) & (3) & (4) \\
\hline 
J1152$+$3400 & $0.39 \pm 0.08$ & $-193 \pm 109$ & $-388 \pm 22$ \\
J0925$+$1403 & $1.41 \pm 0.28$ & $-135 \pm 35$ & $-438 \pm 19$ \\
J1442$-$0209 & $1.29 \pm 0.26$ & $-250 \pm 24$ & $-533 \pm 12$ \\
J1333$+$6246 & $1.25 \pm 0.25$ & $-311 \pm 54$ & $-624 \pm 40$ \\
J1503$+$3644 & $1.09 \pm 0.22$ & $-128 \pm 33$ & $-367 \pm 20$ \\
Haro~11 & $2.10 \pm 0.42$ & $-208 \pm 8$ & $-688 \pm 20$ \\
J0921$+$4509 & $1.13 \pm 0.06$ & $-105 \pm 15$ & $-474 \pm 19$ \\
\hline
\end{tabular}
\tablefoot{Same as Table~\ref{tab:si2} but for \siiiip~1206\AA. There are two fewer galaxies with \siiii coverage because the detector gap covers the \siiiip~1206\AA\ line for the \citet{Leitherer16} sample.}
\label{tab:si3}
\end{table}

The control and \lya\ sample are observed with the G130M grating. The spectra are downloaded from the MAST server, processed through {\small CALCOS} v2.20.1, and reduced following the methods outlined in \citet{wakker2015}. These spectra are convolved to the resolution of the G140L grating to provide similar resolutions for all spectra. 

\begin{figure*}
\includegraphics[width = \textwidth]{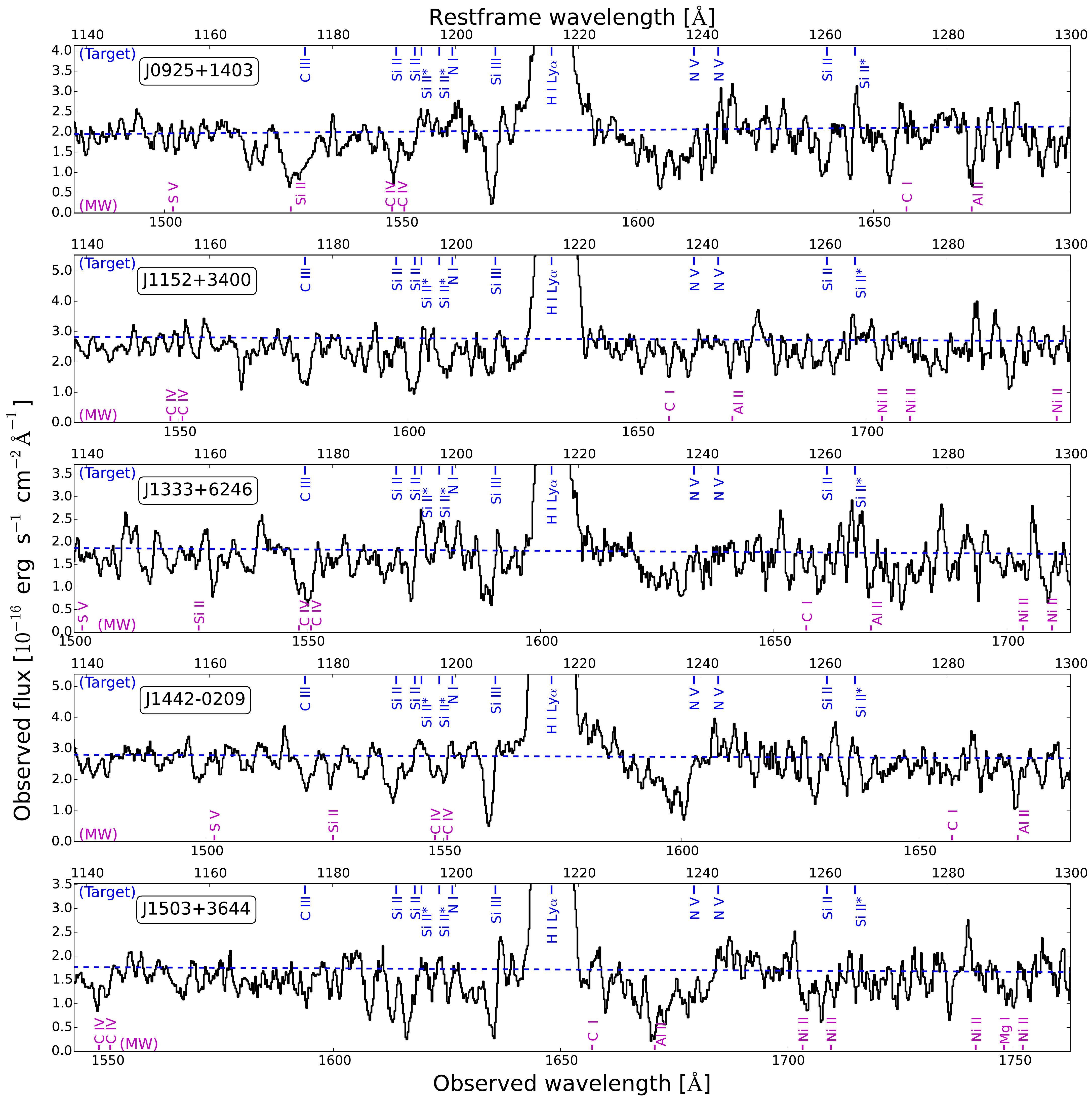}
\caption{Cosmic Origins Spectrograph ultraviolet spectra of the five LyC leakers from \citet{Izotov16, Izotov16b}, with the global first order polynomial fit shown as the dashed blue line. The spectra are ordered from top to bottom as J0925$+$1403, J1152$+$3400, J1333$+$6246, J1442$-$0209 and J1503$+$3644. The observed wavelength is given on the bottom axis and the restframe wavelength is given on the upper axis. Along the upper axis we label interstellar absorption lines in the restframe of the target galaxy, while the lower axis shows possible Milky Way contamination.}
\label{fig:fullspec}
\end{figure*}

All of the control sample, and seven of the nine leakers, have full coverage of both the \siiip~1260\AA\ and \siiiip~1206\AA\ absorption lines. The detector gap covers the \siiiip~1206\AA\ line in the \citet{Leitherer16} sample, therefore we cannot use the \siiii properties for these three galaxies. These silicon lines are two of the strongest tracers of gas in the restframe ultraviolet, and trace partially neutral (\siiip) and fully ionized (\siiiip) gas entrained in photoionized galactic outflows. 
\begin{figure*}
\includegraphics[width = \textwidth]{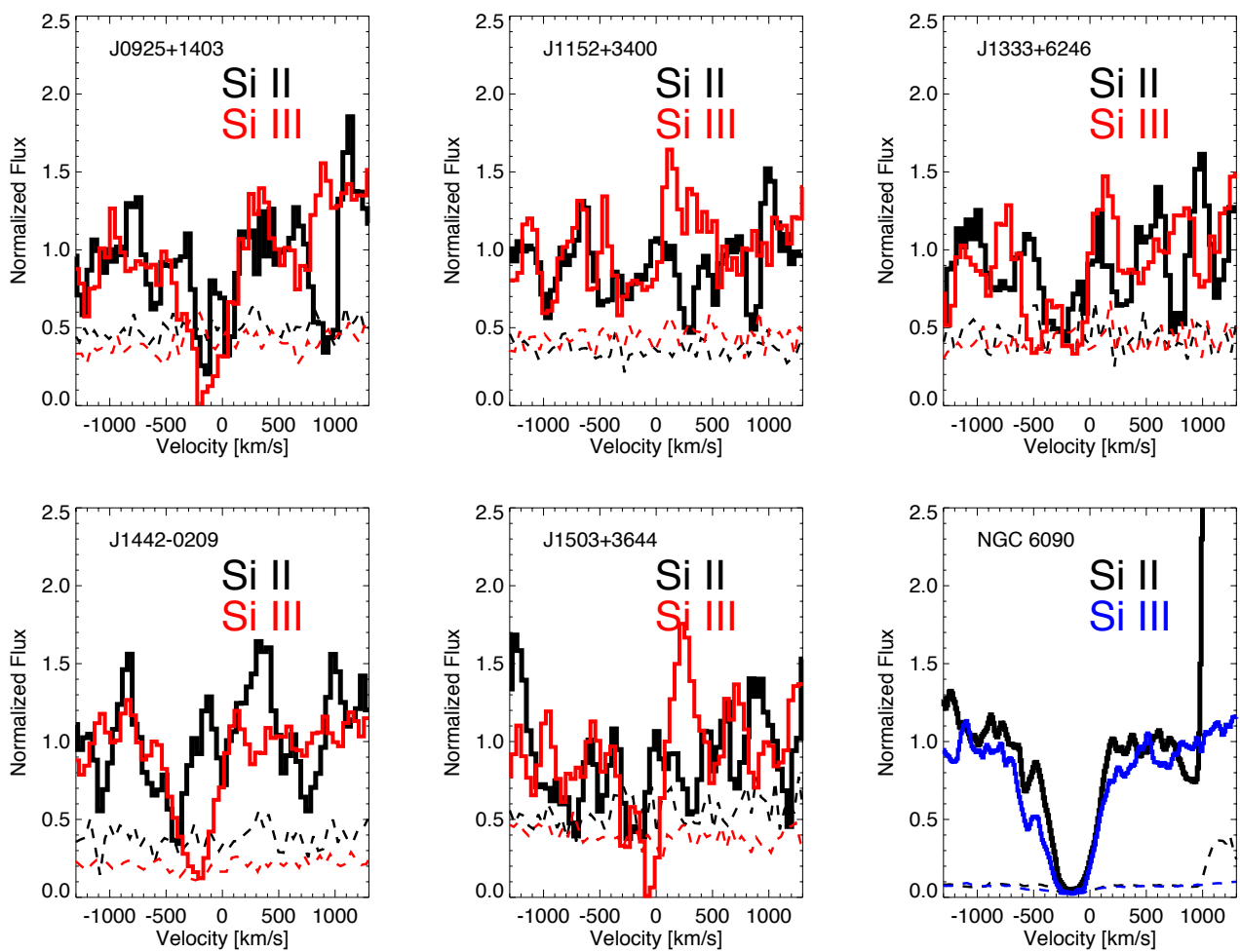}
\caption{Plots of the continuum normalized \siiip~1260\AA\ (black) and \siiii 1206\AA\ (red) absorption profiles. The associated error arrays are given by the dashed lines with colors corresponding to the individual transitions. Each panel is labeled with the name of the galaxy in the upper left. The lower right panel shows the profile from NGC~6090, a galaxy in the control sample, with \siii and \siiii in black and blue, respectively.}
\label{fig:siprofiles}
\end{figure*}

\subsection{Continuum normalization}
The spectra are placed into the restframe using custom fits to emission lines from the Sloan Digital Sky Survey spectra \citep[see Table~\ref{tab:sample};][]{Ahn14}. We set the unity flux level of the leakers by dividing the spectra by a first order polynomial fit to line-free regions, which removes global variations like the spectral slope. We plot the spectra for the five LyC leakers from \citet{Izotov16b} in Fig.~\ref{fig:fullspec}, along with the polynomial fit as the dashed blue line. Prominent absorption lines are marked in Fig.~\ref{fig:fullspec}: lines in the restframe of the target are displayed at the top of each panel, while Milky Way lines are indicated at the bottom. We further normalize the spectra with a local spline fit to only the 20\AA\ around, but excluding the 5\AA\ immediately adjacent to, the \siii and \siiii lines. This accounts for local contributions like broad \lya\ emission or absorption (see the region near \siiii for J0925$+$1403, J1333$+$6246 and J1442$-$0209 in Fig.~\ref{fig:fullspec}).

The control sample is continuum normalized by a simultaneous fit of a {\small STARBURST99} continuum \citep{claus99, claus2010} and a Lorentzian Ly$\alpha$ profile \citep{Chisholm15}. We normalize the control sample differently because the control sample generally has a larger signal-to-noise ratio than the leaker sample, such that we can measure the weak photospheric absorption lines that constrain the stellar population. We set the control sample's final zero-velocity  by cross-correlating the observed flux with the {\small STARBURST99} fit.

\subsection{Outflow properties}

LyC photons ionize neutral gas. Consequently, tracing \hi\ gas is the best way to explore the effect of outflows on LyC escape. With an ionization potential of 16.3~eV, \siiip~1260\AA\ traces partially neutral gas, while the \siiiip~1206\AA\ line, with an ionization potential of 34~eV, traces fully ionized gas. These two lines are ideal for this study because they: are two of the strongest transitions within the restframe UV, are rough approximations of both neutral and ionized gas, and have the same abundances. The \siii and \siiii profiles from the five \citet{Izotov16, Izotov16b} leakers are shown in Fig.~\ref{fig:siprofiles}. In the bottom right panel of Fig.~\ref{fig:siprofiles}, we include the \siiip~1260\AA\ and \siiiip~1206\AA\ profiles from NGC~6090, a large local star-forming galaxy from the control sample. The \siii and \siiii lines from NGC~6090 are stronger and broader (\siii and \siiii equivalent widths of 1.9 and 2.6\AA, respectively) than the profiles from the LyC leakers. Below, we use the kinematics from the absorption profiles to describe the outflow properties. Absorption lines only probe gas between the continuum source and the observer, ensuring that only gas along the line-of-sight to the star-forming region is measured. However, there can be absorption from gas at systemic velocities within the galaxy or the circumgalactic medium. 

We characterize the outflow kinematics using two different velocity measurements: 1) the blue velocity at which the flux reaches 90\% of the continuum level (\vnp) and 2) the velocity at 50\% of the total equivalent width (\vcenp). These two velocity estimates define the outflow in different and complementary ways. \vcen probes the central velocity of the profile, while \vn probes the maximum extent of the absorption. \vn is sensitive to the definition of the continuum level (which we estimate to be uncertain at the 20\% level), while \vcen is sensitive to gas not in the outflow. 

The ratio of the two measurements (\vnp/\vcenp) estimates the asymmetry of the absorption line. A ratio close to 1 implies that more of the absorption is at high velocities, while a high ratio implies that there is a large, low equivalent width tail to shorter wavelengths (see Fig.~\ref{fig:rat_comp} below). Absorption profiles from galactic outflows are typically characterized by a \lq{}{}\lq{}{}saw-tooth\rq{}{}\rq{}{} profile, with a long extend gradual blue wing \citep[see Sect.~\ref{rat};][]{Weiner2009} and a large \vnp/\vcen ratio.

We are also interested in quantifying the strength of the \siii and \siiii transitions. Column density robustly measures the transition strength because it accounts for the integrated number of particles along the entire line-of-sight. However, since we use the lower resolution G140L grating, we do not have sufficient spectral resolution to resolve narrow, high column density features that could comprise the bulk of the column. Further, the \siiip~1260~\AA\ line is strong ($f$-value of 1.22), and saturates at moderate column densities and equivalent widths greater than 1~\AA. Instead, we use the equivalent width of the lines as a proxy of the line strength. Interpreting the equivalent width as the strength of the line is challenging because the equivalent width is triply degenerate with the column density, covering fraction, and line-width. We return to this issue in Sect.~\ref{ew}.

The errors for the velocities and equivalent widths are estimated by varying the observed flux by a random Gaussian kernel centered on zero with a standard deviation equal to the observed error on the flux. We then recalculate the equivalent width, \vcenp, and \vn for this Monte Carlo flux array. We tabulate these results and repeat the procedure 1000 times to produce a distribution of parameter estimates. The standard deviation of this distribution is used as the error for each parameter. The \siiip~1260\AA\ and \siiiip~1206\AA\ values for the leakers are given in Table~\ref{tab:si2} and Table~\ref{tab:si3}, respectively. We calculate the equivalent widths and velocities for the control sample before and after we degrade the spectral resolution, and find that the equivalent width and \vn decrease by a mean of 9 and 4\%. 
\\\
\subsection{\lya\ properties}
\label{lya}
LyC and \lya\ emission are both intimately tied to neutral hydrogen because \hi\ absorbs or scatters the photons. Previous studies found a correlation between the \lya\ and LyC escape fractions \citep[Fig.~2 in][]{verhamme16, dijkstra16}. Additionally, the velocity separation of individual \lya\ emission peaks are related to the \hi\ column densities of the outflows \citep{Kunth98, Dijkstra06, Hashimoto15, verhamme15}. To examine the connection between escaping photons and galactic outflows, we form the \lya\ sample by taking the LARS galaxies from  the control sample and adding Green Peas from \citet{Henry15}. While the escape of \lya\ is a complicated problem, we show in \autoref{lya_prop} that there are \lya\ properties that correlate with \siii properties. \citet{Henry15} measure the maximum velocity differently than we do, and we exclude their \vn points from the \lya\ sample. The \lya\ sample dramatically increases the \hi\ column density dynamic range because \lya\ photons escape at higher column densities than LyC photons.
\\\

We focus on two \lya\ properties: the \lya\ peak velocity separation and the \lya\ escape fraction (\fesclyp). Both of these quantities correlate with the \fesc \citep{verhamme16}. We measure the \lya\ peak velocities as the velocity at which the \lya\ flux reaches a maximum. If there are two peaks -- typically one peak is redward of zero velocity and one peak is blueward of zero velocity -- we measure the velocity of each peak. We then take the velocity separation of these two peaks as the \lya\ peak separation. \fescly is the ratio of two \lya\ to \ha\ flux ratios: (1) the observed ratio, after correcting for internal reddening, and (2) the ratio set by Case B recombination. We take the \ha\ flux values from the SDSS DR1--10 \citep{Ahn14}, which have similar fiber diameters as the 2\farcs5 COS aperture (3\farcs0 for DR1--DR9 and 2\farcs0 for DR10). Many of the leakers are included in DR1--9, but two of the \citet{Izotov16b} leakers were observed with 2\farcs0 fibers because they are from DR10. Finally, since \lya\ is more extended than H$\alpha$, it is impossible to aperture correct the \lya\ in a meaningful way without \lya\ images. Therefore, the values we quote here are uncorrected for aperture effects.
\\\

\subsection{Measuring \fesc}

We use our custom reductions of the archival leakers to reassess previous \fesc values in a consistent way. We measure the mean flux in the rest frame wavelength range between 880-912~\AA\ using a maximum likelihood method. We then compute the statistical Poisson error, and estimate the systematic error from the propagated background subtraction error spectrum \citep{Worseck16, Izotov16b}. In particular, our procedure uses coadditions of dark exposures to estimate the dark current with negligible systematic errors. In addition to the eight LyC leakers detected by COS, we include Haro~11, which has a \fesc measured by FUSE \citep{leitet11, Leitet13}, in the sample as a LyC emitting candidate.  In the Appendix, we compare our remeasured \fesc values to previous values, and explore systematic effects for our \fesc values. 

\section{Results}
\label{results}
Here we compare the outflow properties of the LyC leakers to the control sample of local galactic outflows. This comparison tests whether extreme outflows lead to the escape of ionizing photons.  First, in Sect.~\ref{outflow_prop}, we compare the outflow properties (velocities and equivalent widths) to determine if the two samples have different outflow strengths and velocities. Second, in Sect.~\ref{host_prop}, we explore whether the leakers' outflow properties follow the scaling relations between host galaxy properties (SFR, \mstarp, and metallicity) and outflow properties found in \citet{Chisholm16}. To further examine the connection between LyC and \lya\ escape, we then compare the \lya\ properties and \siii outflow properties in Sect.~\ref{lya_prop}.  Finally, we conclude by exploring the effect of outflows on the LyC escape fractions (Sect.~\ref{fesc}).
\\

\subsection{Comparing outflow properties}
\label{outflow_prop}

\begin{figure*}
\includegraphics[width = \textwidth]{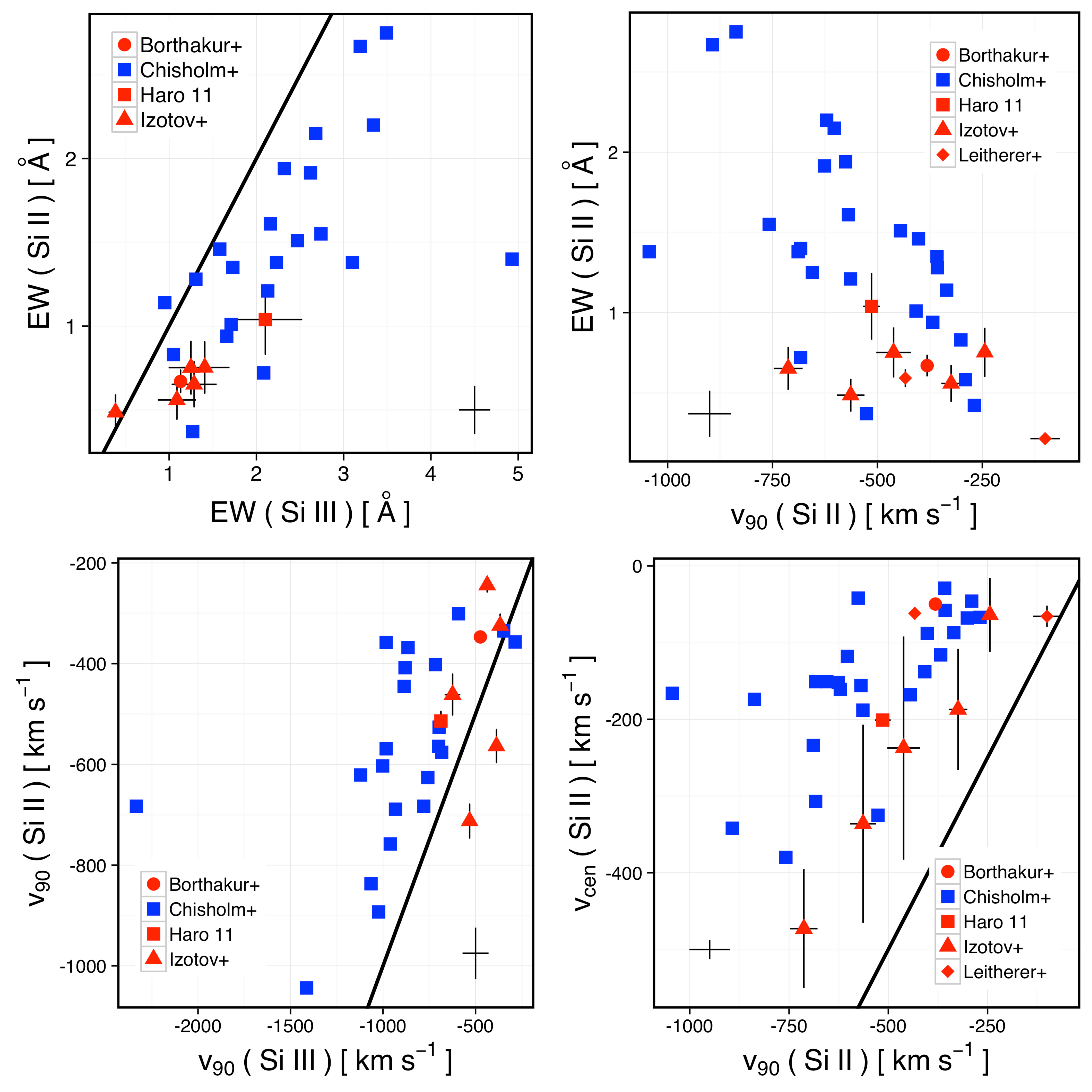}
\caption{Comparison of the outflow properties for LyC leakers (red symbols) and the control sample (blue squares). The upper left panel gives the relationship between the \siii equivalent width and \siiii equivalent width. The upper right panel shows the variation of the \siii equivalent width with the \siii velocity at 90\% of the continuum (\vnp). The bottom left panel compares the \vn of the \siii and \siiii transitions. The bottom right panel compares the equivalent width weighted velocity of the \siii line (\vcenp) to the \siii \vnp. Black lines mark one-to-one relationships. The black crosses in all panels indicate the typical uncertainties for the control sample. Note that we only measure LyC upper-limits for the \citet{Leitherer16} sample (red diamonds; see the Appendix).}
\label{fig:outflow}
\end{figure*}

Fig.~\ref{fig:outflow} compares the outflow properties of LyC leakers (red symbols) to the control sample (blue points). Plots involving \siiii have two fewer leakers because the \citet{Leitherer16} sample has a detector gap over the \siiiip~1206\AA\ transition. In this paper, we focus on the \siiip~1260\AA\ line because the ionization state is similar to \hi, and it provides a rough approximation to neutral gas.

In the upper left panel of Fig.~\ref{fig:outflow} we plot the variation of the  \siii equivalent width with the \siiii equivalent width. The leakers have both low \siii (partially neutral gas) and \siiii (fully ionized gas) equivalent widths, however, they lie on the equivalent width  trend established by the control sample. This trend shows that the \siii equivalent widths are systematically  $\sim40$\% smaller than the \siiii equivalent widths. Eight of the nine leakers have \siii equivalent widths that are less than 1~\AA, and are unlikely to be heavily saturated, even for the strong \siiip~1260\AA\ transition. This emphasizes a defining characteristic of the leaker sample: they have small equivalent widths (discussed further in Sect.~\ref{ew}).

The upper right panel of Fig~\ref{fig:outflow} compares the \siii equivalent width to the \siii \vn (the maximum velocity). At a given equivalent width, the maximum velocities of the leakers are on the upper envelope of the control sample. As discussed below, this is largely because the leakers have smaller equivalent widths, not because they have abnormal \vnp. Additionally, two leakers (J1152$+$3400 and J1442$-$0209) are offset from the trend; although the control sample has three similarly high velocity outliers (J0926, IRAS08339, and GP1054). 

The \siii and \siiii \vn is compared in the lower left panel. For comparison purpose, a one-to-one line is included in black, demonstrating that both samples generally have larger \siiii velocities than \siii velocities. Two of the leakers (J1152$+$3400 and J1442$-$0209, again) have larger \siii \vn at a given \siiii \vnp, suggesting that their neutral gas outflows are at higher velocities than their ionized outflows. However, the other five leakers follow the trends of the control sample. 

\begin{figure*}
\includegraphics[width = \textwidth]{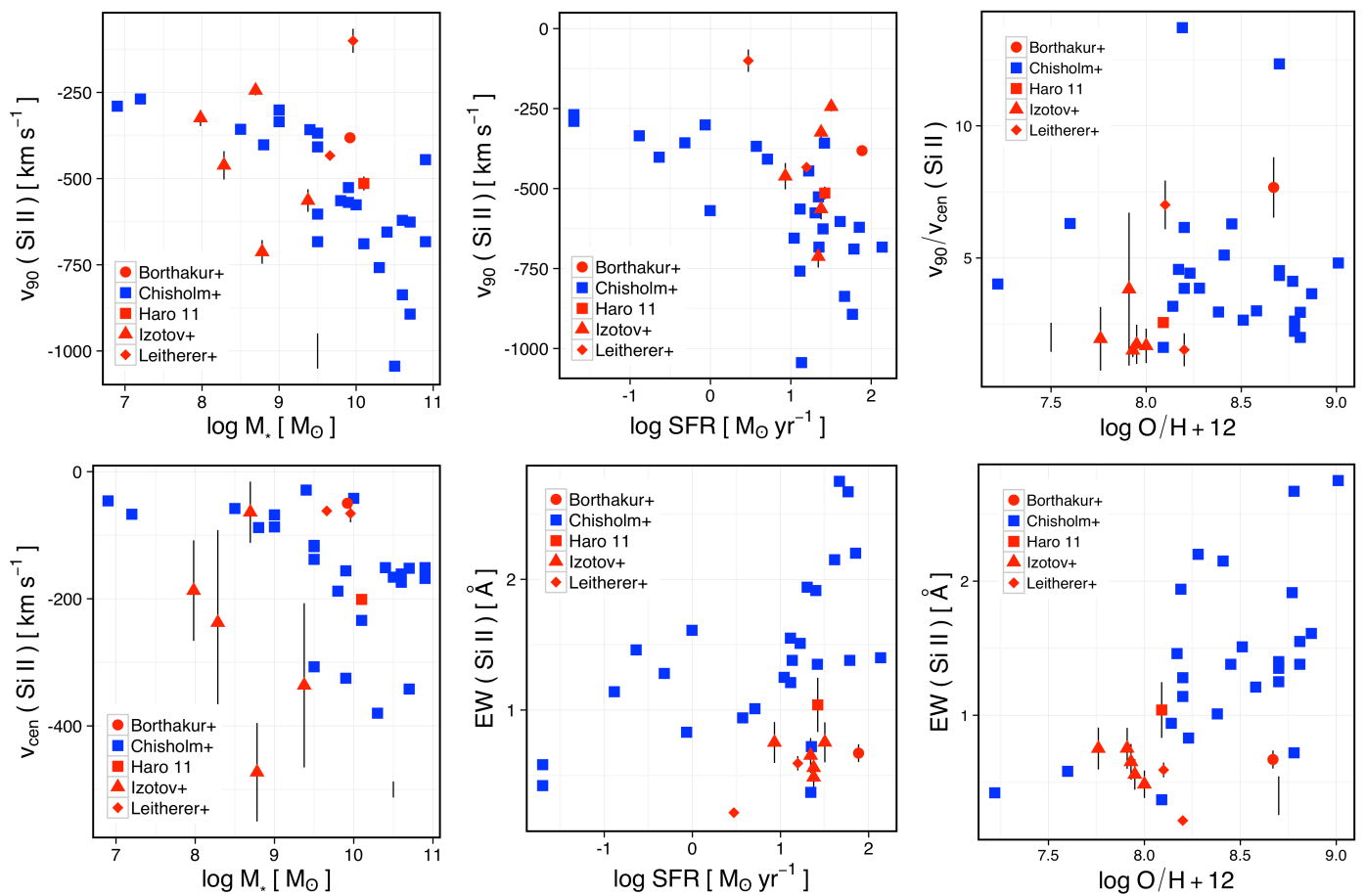}
\caption{Comparison of the outflow properties to their host galaxy properties. The red symbols correspond to confirmed LyC leakers, and the blue points correspond to the control sample. The upper left plot gives the scaling of \siii velocity at 90\% of the continuum (\vnp) with the galaxy's stellar mass (\mstarp). The upper middle panel shows the variation of the \siii \vn with the star formation rate of the galaxy (SFR). The upper right panel shows the variation of the \vnp/\vcen ratio with log(O/H)+12. The lower left panel shows the variation of \vcen with \mstarp. The bottom middle panel shows the variation of the \siii equivalent width with SFR. The lower right panel gives the variation of the \siii equivalent width with log(O/H)+12. The scatter of the velocity and equivalent width depends on which host property is studied. Notably the \vn is more scattered with SFR than with \mstarp; and the equivalent width and SFR are uncorrelated, but the metallicity correlates with the equivalent width. This casually indicates that some host properties more strongly determine outflow properties than others. Note that we only measure LyC upper-limits for the \citet{Leitherer16} sample (red diamonds; see the Appendix).}
\label{fig:host}
\end{figure*}

Finally, the \vcen (a proxy of the central velocity) is compared to \vn in the lower right panel. The five leakers from \citet{Izotov16b} and one from \citet{Leitherer16} (Tol~0440-381) have a \vcen that approaches their \vnp, indicating that most of the leakers profiles are skewed towards blue velocities. The median \vnp/\vcen of the leaker sample is 1.95 as compared to 3.9 for the control sample. The five leakers with lowest \vnp/\vcen ratios also have the highest \fesc values (see Table~\ref{tab:sample}). Meanwhile, the leakers with lower \fesc have \vnp/\vcen ratios more consistent with the control sample. We discuss this further in Sect.~\ref{rat}. 

The Kolmogorov-Smirnov test (KS-test) is a non-parameteric test of whether data-sets are drawn from similar distributions. A hypothesis is made that the two samples are drawn from similar distributions, then the KS statistic tests the statistical significance of the alternative hypothesis that the two samples are drawn from different distributions. The KS-test cannot rule out that the \siii \vcen and \vn are drawn from similar distributions as the control sample  at the 1.5$\sigma$ significance level. Conversely, the KS-test shows that the leaker equivalent widths are drawn from a distribution with a smaller mean equivalent width at a 3$\sigma$ significance (p-value < 0.001). Further, the KS-test finds that the leakers have lower \vnp/\vcen ratios at the 2$\sigma$ significance (p-value < 0.02). 

Since we treat the LyC escape from \citet{Leitherer16} as an upper limit, it is important to ensure that the upper limits do not drive the conclusions about the leakers' outflows. Removing these two upper limits from the leaker sample would increase the median leaker equivalent width by 0.02~\AA; the leakers still have small \siii equivalent widths without the two upper limits. The median of the \vn and \vcen change by $-28$ and $-13$~\kmsp, respectively, when removing the upper limits (6 and 7\% respectively). The similar changes in \vn and \vcen ensures that the \vnp/\vcen ratio does not appreciably change when removing the upper limits. Similarly, without the \citet{Leitherer16} points the leakers still show that they are a drawn from a smaller equivalent width distribution than the control sample: the KS-test statistic decreases from 3$\sigma$ significant to 2.5$\sigma$. This decrease is largely because the sample size is reduced by 25\%. These tests statistically show that the \citet{Leitherer16} upper limits do not drive the observed outflow properties.

We conclude that the \siii outflows, tracing partially neutral gas, are weaker, but reside on the established trends of the control sample. The \siii outflow velocities are marginally larger at fixed equivalent width than the control sample, but are not statistically extreme. Two of the highest \fesc galaxies, J1152$+$3400 and J1442$-$0209, have larger \siii \vn than galaxies at similar equivalent widths. The leakers tend to have lower \vnp/\vcen ratios than the control sample, such that the highest \fesc galaxies have the smallest \vnp/\vcen ratios. These are the major ways that the leakers\rq{} outflow properties differ from the control sample. In the next section, we explore whether these outflow properties are consistent with the relationships between outflows and their host galaxies. This probes whether the small equivalent widths and \vnp/\vcen ratios arise because the leakers have different host or outflow properties. 

\subsection{Comparing host property scaling relations}
\label{host_prop}

In the previous section, we found that leakers have smaller equivalent widths, and central velocities that are closer to their maximum velocities, than the control sample. Here, we explore how the outflow properties depend on the host galaxy properties, to test if the leakers have discrepant host galaxies properties, or different outflow properties from the control sample. Previous outflow studies find that the outflow velocity correlates shallowly with host galaxy properties, but the exact cause of these trends is uncertain: some studies find that outflow velocity scales with SFR \citep{martin2005, rupke2005, Weiner2009, Heckman15, Heckman16}, others find the velocity to scale with \mstar \citep{Weiner2009, martin2012, rubin2014, Heckman15, Heckman16}, and still others find it scales with star formation surface density \citep[\sfrsdp;][]{kornei2012, Heckman15, Alexandroff15, Heckman16}. 

The control sample shows a statistically strong correlation between both the SFR and \mstar of the host galaxies and the velocities and equivalent widths \citep[blue squares in Fig.~\ref{fig:host};][]{Chisholm15, Chisholm16}. Fig.~\ref{fig:host} shows that the leakers' outflow properties scatter more with certain host galaxy properties than with others. Since the leaker sample covers a different portion of the SFR-\mstar plane (see Fig.~\ref{fig:mainseq}), this may suggest that certain host galaxy properties drive the trends in velocity and equivalent width more than others. For example, at a SFR near 10~\sfrp, the outflow velocities spread over nearly the full \vn range, but these velocities relax back onto the \mstar trend (top left panel). A scatter above a SFR of 10~\sfr has been previously seen \citep{martin2005}, and was suggested as a saturation of outflow velocity as ram pressure becomes less efficient at SFRs greater than 10~M$_\odot$~yr$^{-1}$. However, Fig.~\ref{fig:host} suggests that the result may arise from degenerate host galaxy parameters. The fact that the correlation is tighter with \mstar may suggest that \mstar drives the \vn trend. Similarly, the leakers' equivalent widths are appreciably scattered at SFRs greater than 10~M$_\odot$~yr$^{-1}$, but lie on the log(O/H)+12 trend (lower right panel of Fig.~\ref{fig:host}). The scaling between equivalent width and metallicity (a 3.8$\sigma$ significant trend, using a Kendall's $\tau$ test) has a larger spread at high-metallicity, but the trend remains significant at the 2.5$\sigma$ significance when only galaxies greater than log(O/H)+12 of 8 are considered. These differences distinguish which host galaxy properties determine the outflow properties.

To statistically test whether the leakers follow similar scaling relations as the control sample, we use the measured trends from \citet{Chisholm16} to define expected outflow properties of the leakers given their SFR, \mstarp, and log(O/H)+12. If the median of the leakers outflow properties are further from the relations than the scatter of the trends, then the leakers produce different outflows, at a given host galaxy property, than the control sample. The \vcen and \vn are consistent, within the scatter of the trend, for \mstar and SFR. However, the equivalent width of the leakers are 0.5~\AA\ outside the scatter of the SFR trend, implying that the leakers have smaller \siii equivalent widths at a given SFR than the control sample. Conversely, at a given metallicity, leakers have equivalent widths consistent with the control sample (lower right panel of Fig.~\ref{fig:host}); metallicity drives the \siii equivalent width trend. These trends are actually strengthened by removing the \citet{Leitherer16} upper limits because those galaxies fall further outside of the scatter than many of the other samples. This demonstrates that the outflows from LyC leakers are largely consistent with the scaling relations established by the control sample, with the equivalent widths largely set by the metallicities of the galaxies.

As mentioned in Sect.~\ref{outflow_prop}, the most notable kinematic difference between the leakers and the control sample is that their \vnp/\vcen ratios are lower. The \vnp/\vcen ratio is uncorrelated with host galaxy properties, displaying a flat distribution across metallicity (upper right panel of Fig.~\ref{fig:host}). We explore possible causes of low \vnp/\vcen ratios in Sect.~\ref{rat}.
\begin{figure*}
\begin{subfigure}{0.33\textwidth}
\includegraphics[width = \textwidth]{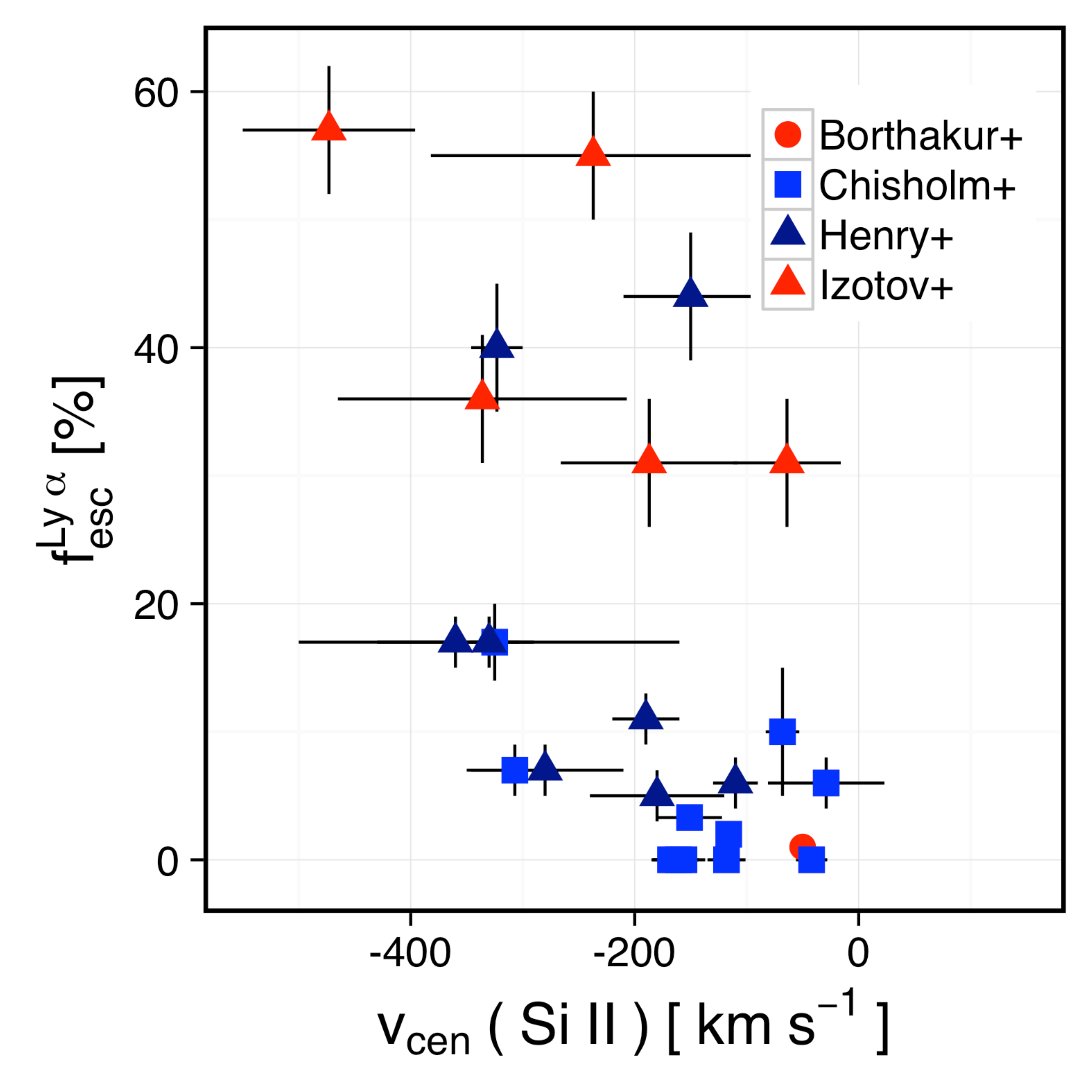}
\end{subfigure}
\begin{subfigure}{0.33\textwidth}
\includegraphics[width = \textwidth]{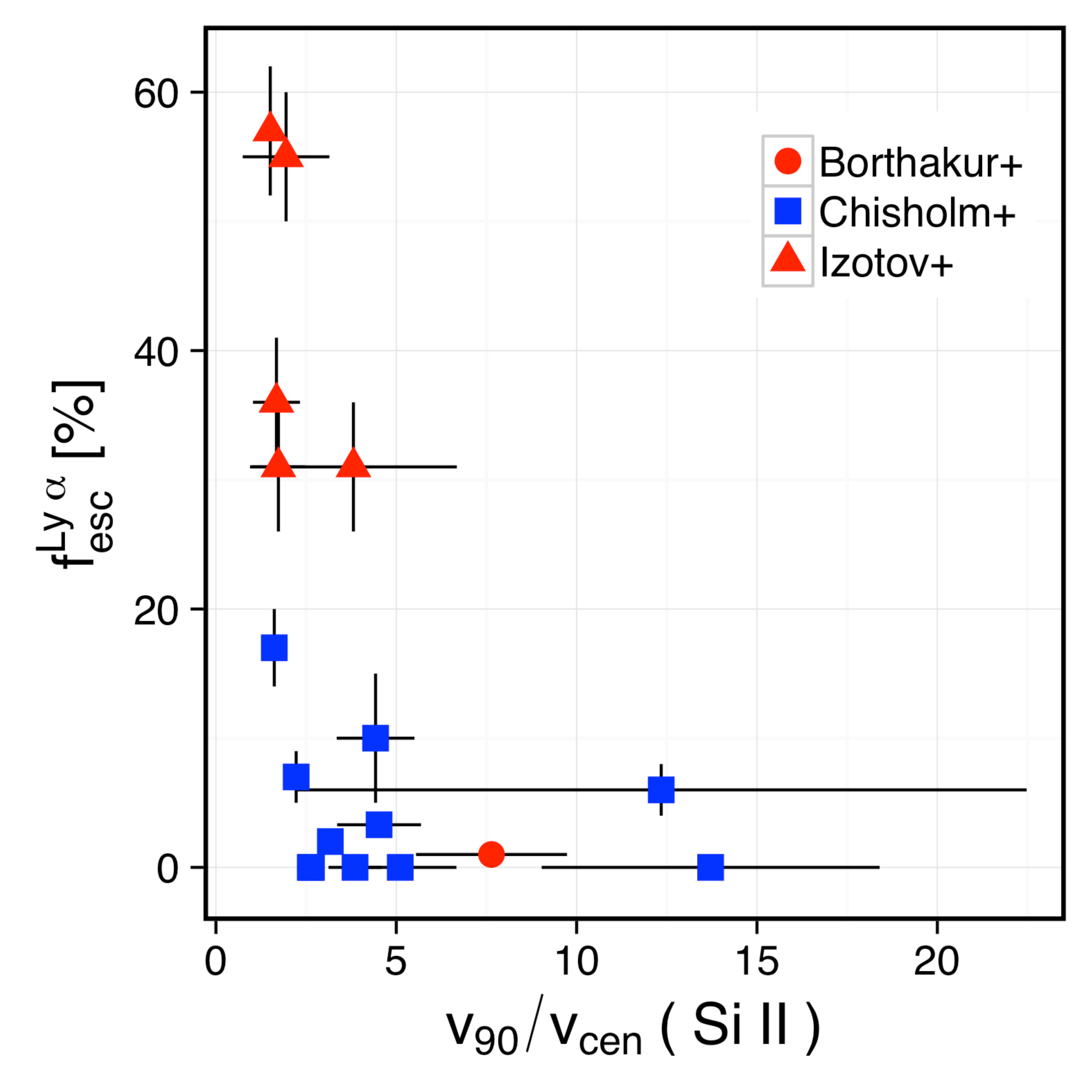}
\end{subfigure}
\begin{subfigure}{0.33\textwidth}
\includegraphics[width=\textwidth]{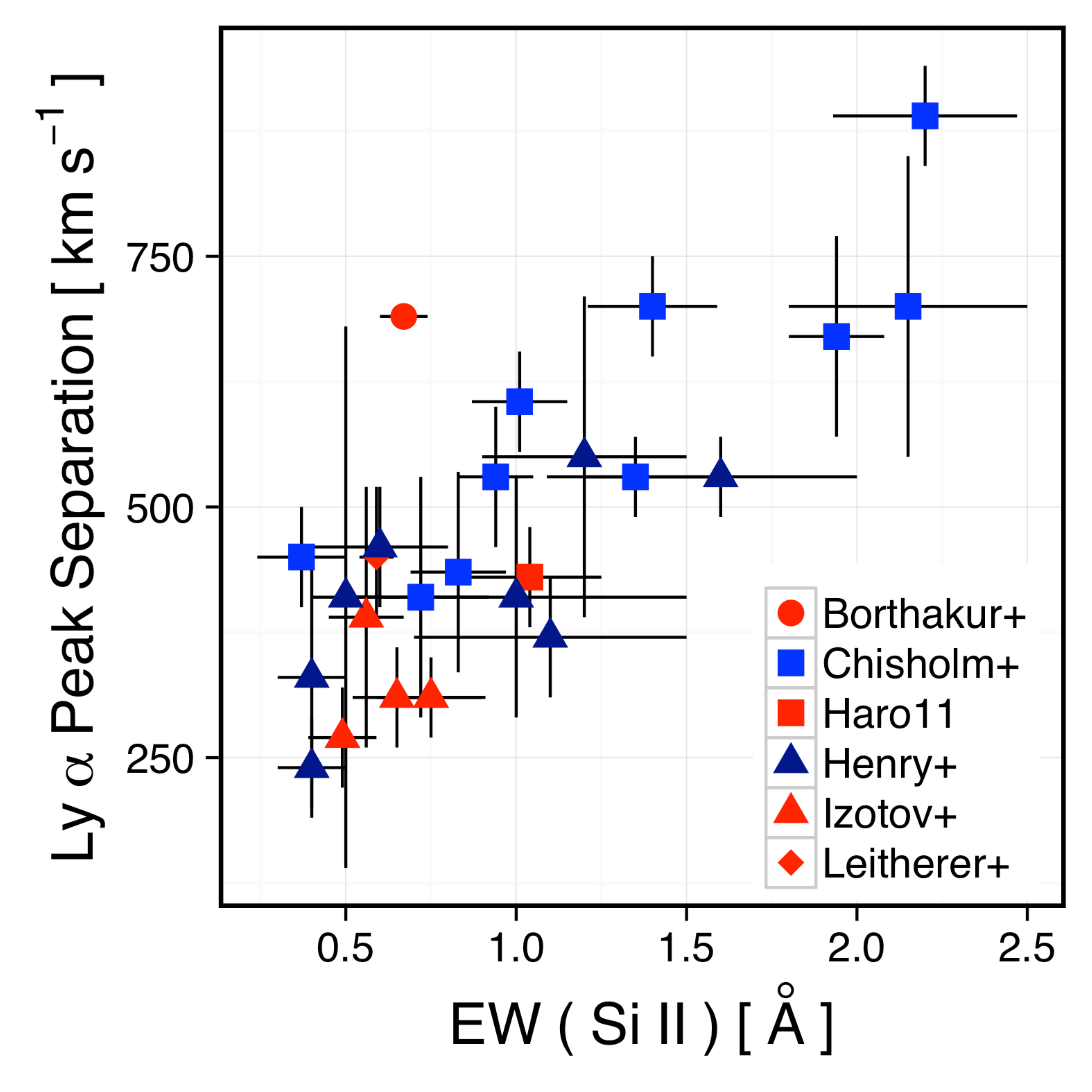}
\end{subfigure}
\begin{subfigure}{0.33\textwidth}
\includegraphics[width = \textwidth]{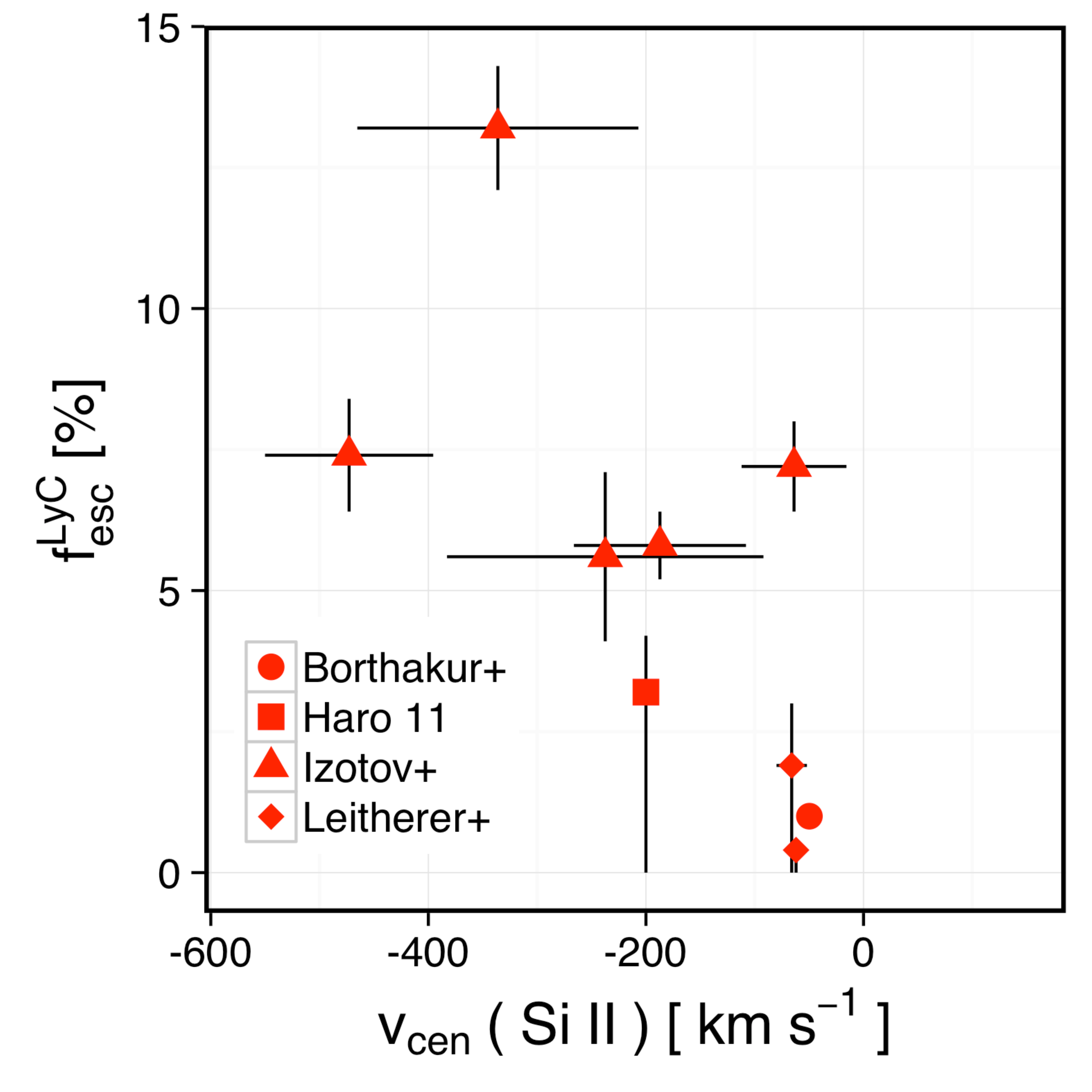}
\end{subfigure}
\begin{subfigure}{0.33\textwidth}
\includegraphics[width = \textwidth]{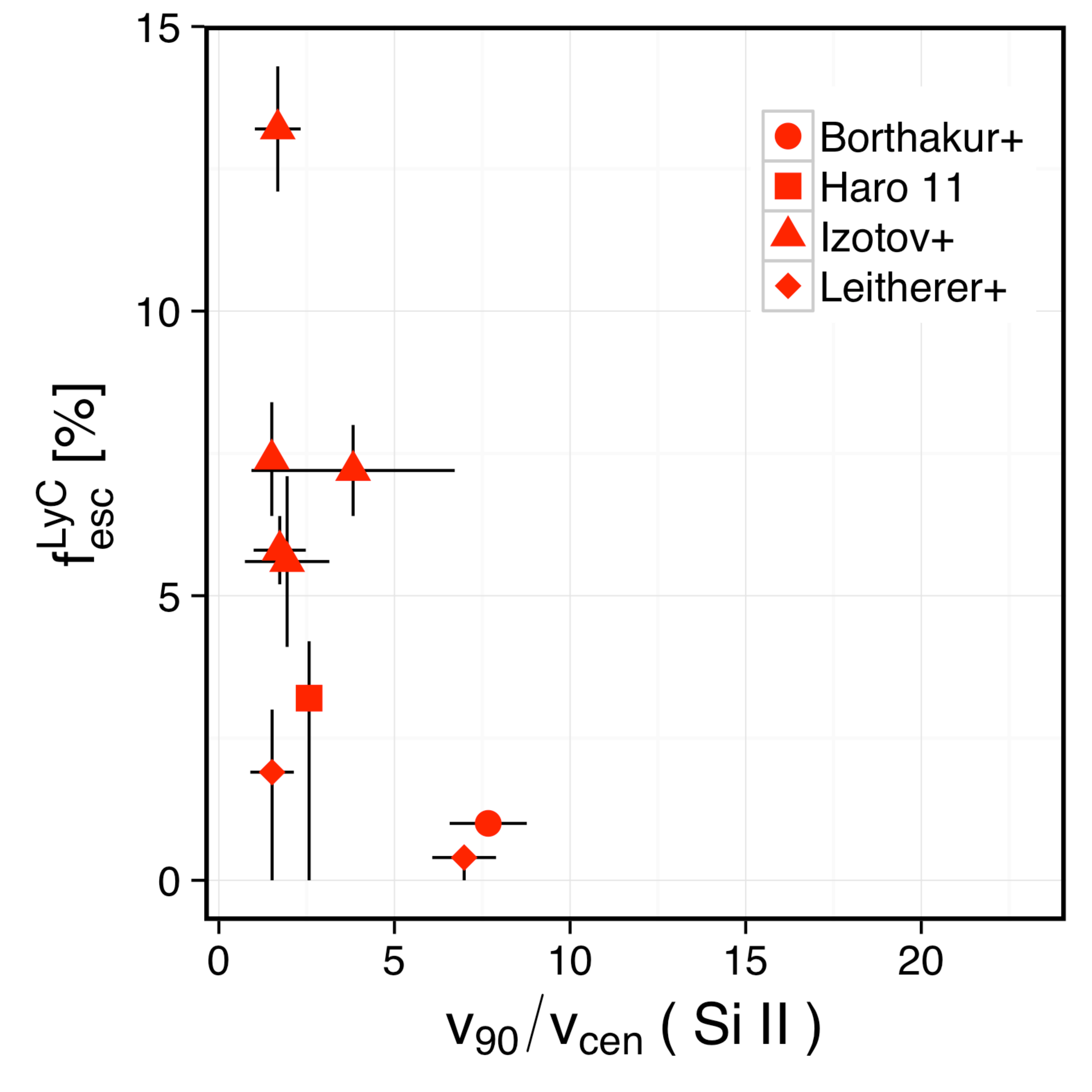}
\end{subfigure}
\begin{subfigure}{0.33\textwidth}        
\includegraphics[width= \textwidth]{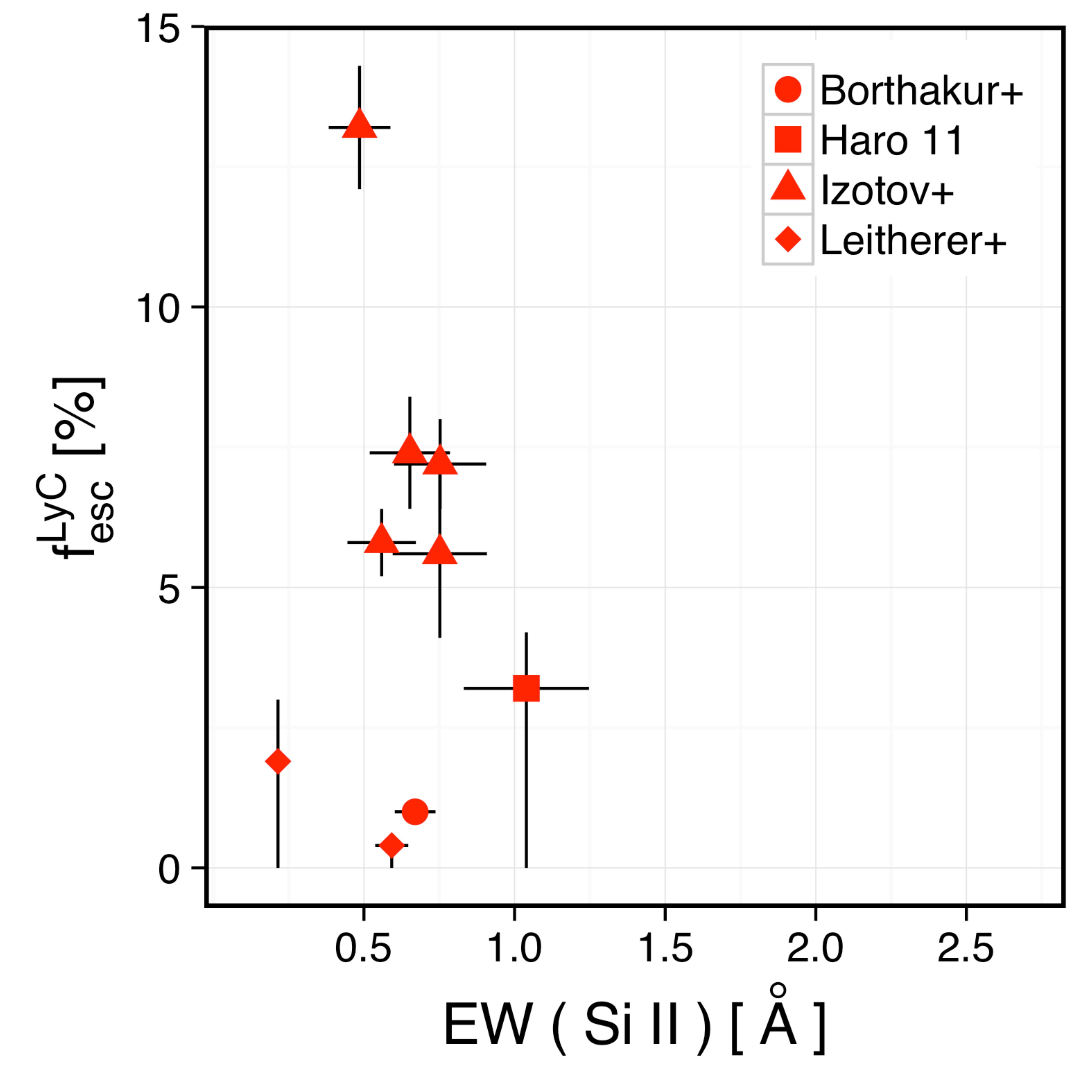}
\end{subfigure}
\caption{Plots of the \lya\ (top panels) and Lyman continuum (lower panels) escape fractions with \siii outflow properties. The left column shows the effect of the \siii \vcen on the escape fractions. The middle column shows the effect of the \vnp/\vcen ratio (a measure of the shape of the line profile) on the respective escape fractions. Finally, the right column shows the effect of the \siii equivalent width on the \lya\ peak separation (upper right panel) and the LyC escape fraction (upper left panel). Both the peak separation and \fesc are related to the \hi\ column density. The point shapes correspond to their respective samples, as shown in the legend.}
\label{fig:fesc}
\end{figure*}

\subsection{Outflows' influence on \lya\ properties}
\label{lya_prop}

\hi\ column densities and outflow velocities control the escape of \lya\ photons \citep{Kunth98, verhamme15}. Here, we use the larger \lya\ sample to explore how the Ly$\alpha$ properties relate to the \siii outflow properties.

The top three panels of Fig.~\ref{fig:fesc} show the variation of \lya\ with \siii outflow properties. In the top left panel we show the Ly$\alpha$ escape fraction (\fesclyp) plotted versus the \siii \vcenp. At low \fesclyp, there is a sequence of increasing \fescly with increasing \siii outflow velocity, but the sequence of increasing \fescly with increasing \siii velocity disappears at higher \fesclyp. The scattered galaxies at large \fescly have the largest \fescp, and the lowest \hi\ column densities. This trend may show that there are two types of \lya\ emitters: 1) high \hi\ column density outflows where the low \lya\ escape depends on the velocity of the scattering medium (the lower sequence) and 2) low density outflows where the \lya\ photons propagate relatively freely through the outflow (upper sequence). Another possible explanation is that \fescly is independent of the outflow velocity, and we poorly sample the full distribution.

The upper middle panel of Fig.~\ref{fig:fesc} shows the effect of the \vnp/\vcen ratio on \fesclyp. High \vnp/\vcen ratios produce small \fesclyp, whereas low \vnp/\vcen ratios occupy the entire range of \fesclyp. When the \vnp/\vcen ratio is high, the bulk of the \hi\ is at low velocities, and it cannot scatter the \lya\ photons in velocity-space enough for the photons to escape the outflow. Meanwhile, when \vcen is closer to the maximum velocity -- when \vnp/\vcen is small -- the bulk of the absorption is at high velocities, allowing the \hi\ to efficiently scatter the \lya\ to higher velocities. Once the kinematics enable \hi\ to scatter \lya\ photons to high velocities, the \hi\ column density and geometry determine if photons escape. The dependency on geometry and \hi\ column density produces the large spread of \fescly seen at low \vnp/\vcenp. This emphasizes outflow kinematics and velocity distributions for \lya\ escape. This point is discussed further in Sect.~\ref{rat}. 

The final \lya\ relation that we explore is between the \lya\ peak separation and the \siii equivalent width (upper right panel of Fig.~\ref{fig:fesc}). The Kendall's $\tau$ correlation test suggests that there is a correlation between these variables at the 3$\sigma$ significance level (p-value less than 0.001; Kendall's $\tau$ value of 0.52). Since the peak separation is correlated with \hi\ column density \citep{Dijkstra06, verhamme15},  the \siii equivalent width is likely correlated with \hi\ column density. We discuss this correlation further in Sect.~\ref{ew}.

\subsection{Outflows' influence on \fesc}
\label{fesc}

With the sample of nine LyC leaking galaxies we test whether the outflow properties (velocities and equivalent widths) lead to higher escape fractions of ionizing photons. The lower left panel of Fig.~\ref{fig:fesc} shows how the \fesc  changes with \siii \vcenp. The galaxies with \fesc greater than 5\% span the full range of \siii \vcenp, while galaxies with small \fesc are clustered at relatively low velocities. The \fesc plot follows the \fescly plot above it, where high \fesc galaxies (i.e. low \hi\ column densities) scatter over the full velocity range, and low \fesc galaxies follow a tighter trend. 
\\\

While the outflow kinematics should affect the \lya\ escape, kinematics do not necessarily influence LyC escape because the photons are not scattered by the outflow. However, in the bottom middle panel of Fig.~\ref{fig:fesc} we show that the \fesc has a similar relationship with the \vnp/\vcen ratio as \fescly does: at low \vnp/\vcen \fesc spans the full range of \fesc values, while at high \vnp/\vcen ratios few LyC photons escape the galaxy. This implies that the kinematics of the outflow influence the number of ionizing photons that escape the galaxy (see Sect.~\ref{rat} below).
\\\

Finally, in the lower right panel of Fig.~\ref{fig:fesc} we show the relationship between \fesc and the \siii equivalent width. While there is not a clear trend with this data, the plot mimics the inverse of the \lya\ peak separation versus \siii equivalent width plot, although is has only a faction of the dynamic range (see the red points in the plot above). This possibly points to a similar origin of LyC and \lya\ escape \citep{verhamme16}, suggests that small \siii equivalent widths might indicate measurable LyC escape fractions.

\section{Discussion}
\label{discuss}

To test whether Lyman continuum leaking galaxies have extreme outflows, we compare the leakers' galactic outflows to a control sample of nearby star-forming galaxies. Generally, we find that the velocities and equivalent widths from leakers' galactic outflows are not larger than those from the control sample. The velocities and equivalent widths lie upon the defined relations of the control sample. However, we do find two intriguing differences: 1) the leakers have smaller \siii absorption line equivalent widths and 2) the ratio of the maximum velocity to the central velocity is lower in the leaker sample than the control sample. Additionally, we find that the \siii equivalent width scales strongly with the \lya\ peak separation, which has been shown to theoretically scale with the \hi\ column density \citep{verhamme15}. These results suggest that the equivalent widths from leakers are smaller than from the control sample, implying that the leakers might actually have {\it less} extreme \siii mass outflow rates than control sample galaxies at similar star formation rates.  Here, we explore how these two differences relate to previous results and high redshift observations. We then discuss the possible implications of the small equivalent widths and \vnp/\vcen ratios.

\subsection{Comparing to previous results}
\label{comp}
\begin{figure*}
\includegraphics[width =\textwidth]{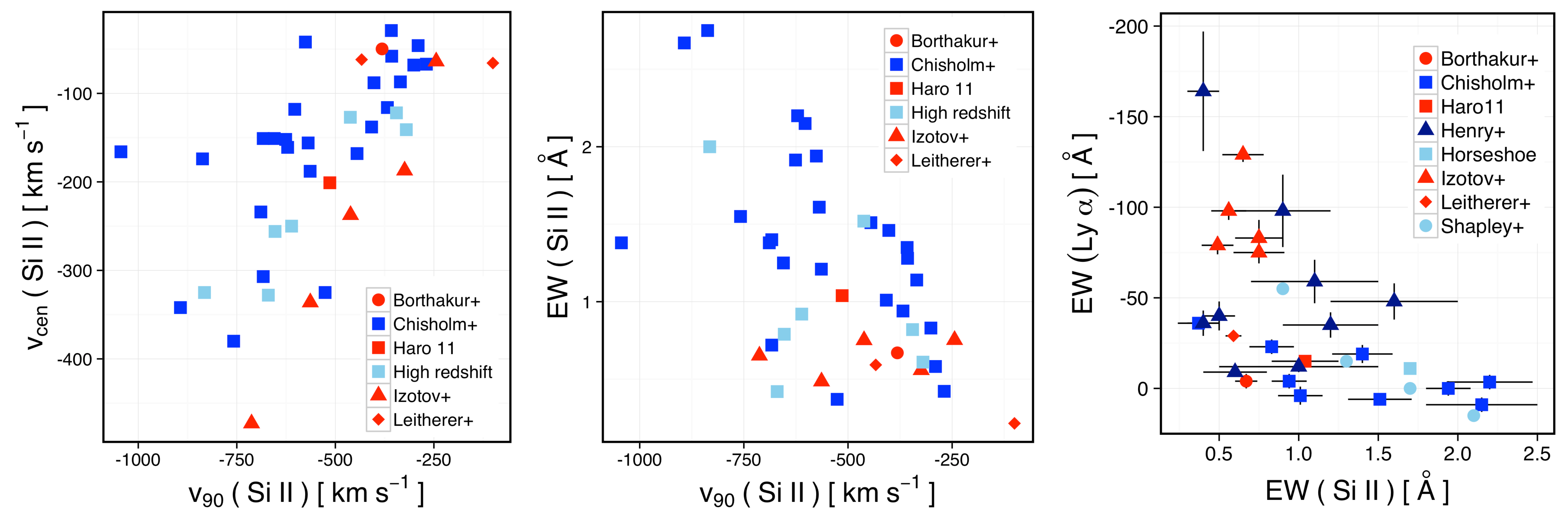}
\caption{The two panels on the left are similar to the right two panels of Fig.~\ref{fig:outflow}, but they include seven high redshift (z$\sim 4$) galaxies from \citet{Jones13} and \citet{leethochawalit} in light blue. These seven high redshift galaxies are LyC candidates from their small \siii covering fractions, and they have similar outflow properties as the control sample. Abell~2390~H5b is an intriguing candidate because it has a lower \vnp/\vcen ratio (2.0)  and low \siii equivalent width. The right panel compares the \lya\ equivalent width to the \siii equivalent width.  Plotted as light blue circles are points from \citet{Shapley03} who stacked $z\sim3$ galaxies based on their \lya\ equivalent widths. The low redshift galaxies encompass the \citet{Shapley03} points, but the leakers are systematically at lower \siii equivalent widths. Large \siii equivalent widths correspond to low \lya\ emission, while low \siii equivalent width span a large range of \lya\ emission properties. Additionally, we include the high redshift galaxy, the Cosmic Horseshoe, as a light blue square. This galaxy is not detected in LyC emission, and correspondingly has a large \siii and small \lya\ equivalent width. Here, negative equivalent widths correspond to emission features.}
\label{fig:highz}
\end{figure*}

While this sample is the largest sample of confirmed low redshift leakers, previous studies have used indirect measurements to infer LyC leakage. \citet{Alexandroff15} use the residual intensity of \siii lines, \lya\ profiles, and optical emission lines to rank-order galaxies in terms of possible LyC leakage, finding that the star formation surface density (\sfrsdp; SFR/Area) and the outflow velocity are the most correlated to possible LyC escape (although only at the 1.6 and 1.8$\sigma$ significance level, respectively). In Fig.~\ref{fig:host} we show that the outflow properties of the confirmed leakers are not statistically different from the control sample, implying that leakers do not generate significantly more extreme outflows than galaxies in the control sample with similar \mstar and SFR. Since we do not know the \fesc properties of the control sample, this does not mean that SFR does not play an important role in LyC escape, but it does mean that the leakers\rq{} outflows are not statistically more extreme than the control sample. \citet{Alexandroff15} find \sfrsd correlates best with leakiness, therefore we investigate the effect of the size (or compactness) of the star-forming region on the outflow velocities using the values from \citet{Alexandroff15} and the half-light radii measured here, but we do not find statistically significant relations between the compactness and the outflow velocity, equivalent width, or \fescp. We conclude that the leakers follow the observed trend between velocity and SFR from the control sample, and that extreme outflows are not required for LyC escape. 
\\\

\subsection{Comparing to high redshift galaxies}
\label{highz}
While low redshift galaxies are analogs to galaxies that could reionize the early universe, it is important to also observe higher redshift galaxies to determine the outflow properties of leakers near the epoch of reionization. However, known low redshift leakers currently outnumber high redshift leakers, making it more challenging to discern the properties of high redshift leakers. For example,  {\it Ion2}, a confirmed high redshift LyC leaker, has a high \fesc of 64\% \citep{deBarros16}, and an undetected \siiip~1260\AA\ line, with an equivalent width upper limit of 0.7\AA. This non-detection is consistent with the small equivalent widths of the leaker sample found here. Further, {\it Ion2} has a strong double peaked \lya\ profile, indicating that there is little \hi\ along the line-of-sight \citep{deBarros16}. Meanwhile, Q1549-C25, a recently discovered $z = 3.2$ leaker, has a weaker [\ion{O}{iii}] emission line, prominent \siii absorption features, and a single-peaked \lya\ profile offset by $+250$~\kms\ relative to the systemic velocity \citep{shapley16}. All of these traits indicate that there is a large \hi\ column density along the line-of-sight, but, surprisingly, \citet{shapley16} measure an \fesc greater than 51\%.

While detections of high redshift leakers provide valuable information about LyC emission, the non-detections also improve our understanding of LyC escape. A2218-Flanking is a recently discovered leaker at $z=2.5$ that is in a group with more massive galaxies \citep{bian}. Strangely, the authors do not detect LyC leakage from the more massive galaxies that have similar star formation rate surface densities, suggesting that stellar mass -- not simply star formation rate surface density -- plays an important role in the escape of ionizing photons \citep{bian}. Unfortunately, restframe UV observations of the \siii metal absorption lines are not available for A2218-Flanking.

We can also use the seven $z\sim4$ lensed galaxies that \citet{Jones13} and \citet{leethochawalit} suggest could be LyC leakers. We measure \vnp, \vcenp, and  equivalent width from their DEIMOS spectra, with resolution of 70~\kmsp, in the same way that we do for the leaker sample. These seven galaxies are suggested to have \fesc between 10--67\% from the \siii covering fractions, but the galaxies do not have confirmed LyC emission. The galaxies have similar SFRs to the leakers in our sample, with values between 7 and 49~\sfrp, after converting to a Chabrier IMF. 

In the left panel of Fig.~\ref{fig:highz} we plot the \siii \vn versus the \siii \vcen of the seven high redshift galaxies in light blue. These high redshift galaxies have similar outflow velocities to the control sample. One galaxy, Abell~2390~H5b, has a \vnp/\vcen ratio of 2.0, but the rest of the galaxies have ratios near 2.5. Additionally, in the middle panel of Fig.~\ref{fig:highz} we show the \siii equivalent width plotted against the \siii \vnp. Similar to the low redshift leakers, the high redshift leaker candidates generally have low equivalent widths and slightly elevated \siii \vnp. The $z\sim4$ leaker candidates do not have significantly different \siii properties from the low redshift leakers, suggesting that the outflow properties do not change appreciably from high redshifts.  

High redshift galaxies also enable a comparison of the \lya\ emission properties to the \siii properties. Using a sample of 118 Lyman Break Galaxies at redshifts of 3, \citet{Shapley03} stack spectra based on their \lya\ equivalent widths, finding a correlation between \lya\ equivalent width and \siii equivalent width (see the light blue circles in the right panel of Fig.~\ref{fig:highz}). For comparison, we measure the \lya\ equivalent width from our \lya\ sample. The low redshift points surround the \citet{Shapley03} stacks and extend the range to higher \lya\ equivalent widths, but they also add enough scatter to obscure the previous trend. Further, at low \siii equivalent widths the \lya\ equivalent width spans the whole range. This is likely because even at low \hi\ column densities the \lya\ escape still depends on dust content, geometry, and outflow properties; ensuring that low column density does not guarantee that \lya\ can escape. Conversely, at high \siii equivalent widths \lya\ photons cannot escape because the large \hi\ column density efficiently absorbs the photons. Further, we include observations from \citet{quider} of the Cosmic Horseshoe, a galaxy that does not leak LyC \citep{vasei}, as a light blue square. Interestingly, this non-leaker has larger \siii and smaller \lya\ equivalent width than the leakers. This agrees with our finding that LyC leakers have small \siii equivalent widths. Leaking from high-redshift galaxies is still unclear, but more high-resolution observations of individual metal lines, coupled with clear LyC detections, will help clarify the situation.

\subsection{Why do leakers have small equivalent widths?}
\label{ew}

A notable property of the LyC leakers is that they have small \siii and \siiii equivalent widths (Fig.~\ref{fig:outflow} and Fig.~\ref{fig:host}). Equivalent widths are challenging to interpret because they are degenerate with a variety of parameters: line-width, column density and covering fraction. Below we walk-through three different mechanisms for decreasing the equivalent width: covering fraction,  metallicity, and \hi\ column density. We conclude that the small equivalent widths are likely set by a combination of small \hi\ column densities, and metallicities. 

As discussed in Sect.~\ref{comp}, one possible mechanism for creating the low equivalent widths is to reduce the covering fraction of the outflowing gas by evacuating the gas from the galaxy. Additionally, \hi\ gas may reside in small clumps within the ionized medium, and this patchy geometry may enable ionizing photons to escape the galaxy relatively easily through low density channels \citep{dijkstra16, gronk16, rivera17}.  Unfortunately, the relatively low signal-to-noise ratio and the degraded spectral resolution of the leaker sample makes measuring the covering fraction impossible. Consequently, we cannot rule out substantial covering fraction variations with the equivalent width. However, \citet{Chisholm16} uses the relatively weak \siiv doublet to show that there is not substantial variation in the covering fraction at line center with equivalent width of the control sample. Since the leakers lie along similar equivalent width distributions as the control sample, this suggests that the leakers\rq{} covering fractions may not lead to small equivalent widths.  However, we stress that at our resolution and signal-to-noise ratio we can neither confirm nor rule out the role that covering fraction plays in establishing the equivalent widths. Further studies at higher resolution and signal-to-noise ratios are required to rule out covering fraction effects (see \autoref{future}).

The leakers' equivalent widths lie along the trend set by the log(O/H)+12 of the galaxy, but scatter with the SFR, indicating that metallicity and the \siii equivalent width are causally related (lower right panel of Fig.~\ref{fig:host}). The metallicity is a natural explanation for changes in silicon equivalent width: the silicon density is equal to the product of the metallicity and the total gas density. Reducing the metallicity reduces the total amount of silicon. While this is the most apparent way that metallicity changes the silicon equivalent width, it also affects \siii in a subtler way: with fewer metals there is less cooling and the temperature of the gas increases. This reduces the neutral gas fraction and shifts the ionization structure of the outflow. This is consistent with the leaker equivalent widths following the control sample because the \siii equivalent width of the control sample also decreases at lower metallicities.

A final way to decrease the \siii  equivalent width is to change the total \hi\ column density. The low level of \hi\ absorption underneath the strong \lya\ emission in Fig.~\ref{fig:fullspec} indicates that there is little \hi\ along the line-of-sight \citep[see also the discussion on the strength of the red wing of \lya\ in][]{reddy}. Typically, local galaxies show deep \lya\ absorption features \citep{Wofford13, Rivera15}, but the leakers only show \lya\ emission, indicating that there is negligible \hi\ along the line-of-sight. This observation is independent of radiative transfer modelling, but radiative transfer simulations provide supplemental information about the \hi\ column density. As discussed in Sect.~\ref{lya_prop}, the \lya\ peak separation strongly correlates with \hi\ column density \citep{Dijkstra06, verhamme15}, where peak separations below 300~\kms\  usually indicate LyC leakage \citep{verhamme16}. The strong correlation between the peak separation and the \siii equivalent width in Fig.~\ref{fig:fesc} suggests that the small \siii equivalent widths are likely due to small \hi\ column densities. Recent simulations of the \lya-LyC connection using clumpy outflows confirm that \lya\ emission traces the LyC leakage \citep{gronk16,dijkstra16}, although only idealized spherical geometries have been investigated, and more sophisticated models are needed to determine the origin of the correlation between \lya\ peak separation and \siii equivalent width.

After exploring our observations, we conclude that the small \siii equivalent widths are likely produced by a combination of decreasing metallicity and decreasing \hi\ column density, but decreasing covering fractions cannot be ruled out due to our low spectral resolution. These observations are more consistent with the density-bounded scenario where the gas density truncates before the neutral hydrogen absorbs the ionizing photons \citep{jaskot13, Nakajima14}. The low \hi\ column densities and metallicities may also explain why the O$_{32}$ ratio efficiently selects LyC leakers because high O$_{32}$ galaxies are typically young, low metallicity systems \citep{Nakajima14, Stasinska15, Izotov16}. This has important consequences for the early universe, where galaxies generally have fewer metals, and therefore could have a larger \fescp. We now turn to the second distinguishing parameter of the leakers' outflows: the \vnp/\vcen ratio. 

\subsection{Do small \vnp/\vcen ratios help LyC photons escape?}
\label{rat}

The second notable trait of the leakers is that they have smaller \vnp/\vcen ratios (median value of 1.9) relative to the control sample (median value of 3.9). Here we explore the physical mechanisms that could create asymmetric line profiles associated with these small ratios. 

The first possibility is that the control sample has a substantial amount of zero-velocity absorption \citep[see Fig. 6 of][]{Chisholm15}, and that the leakers do not. The large zero-velocity absorption drives the \vnp/\vcen ratio to values higher than 4. The zero-velocity medium may absorb most of the ionizing photons,  but the leakers do not have this high column density static medium. To be general, for the rest of the discussion of \vnp/\vcenp, we neglect the effect of zero-velocity absorption, but this could drive LyC escape.

Next, we turn to the line profiles of galactic outflows. The line profile is determined by how the optical depth and covering fractions vary with velocity. Under the Sobolev approximation \citep{sobolev}, the optical depth of an expanding medium is determined by the density and velocity gradient of the outflow, such that
\begin{equation}
\tau(v) =  \tau_0  \mathrm{n}(v) \frac{\mathrm{dr}}{\mathrm{d}v}
\label{eq:sobop}
\end{equation}
where $\tau_0$ is a constant that depends on the atomic physics of the transition, n($v$) is the density at a given velocity ($v$), and dr/d$v$ is the radial velocity gradient of the outflow. Consequently, the acceleration of the outflow (dr/d$v$) controls the shape of the line profile.

\begin{figure}
\begin{subfigure}{\textwidth}
\includegraphics[width = 0.5\textwidth]{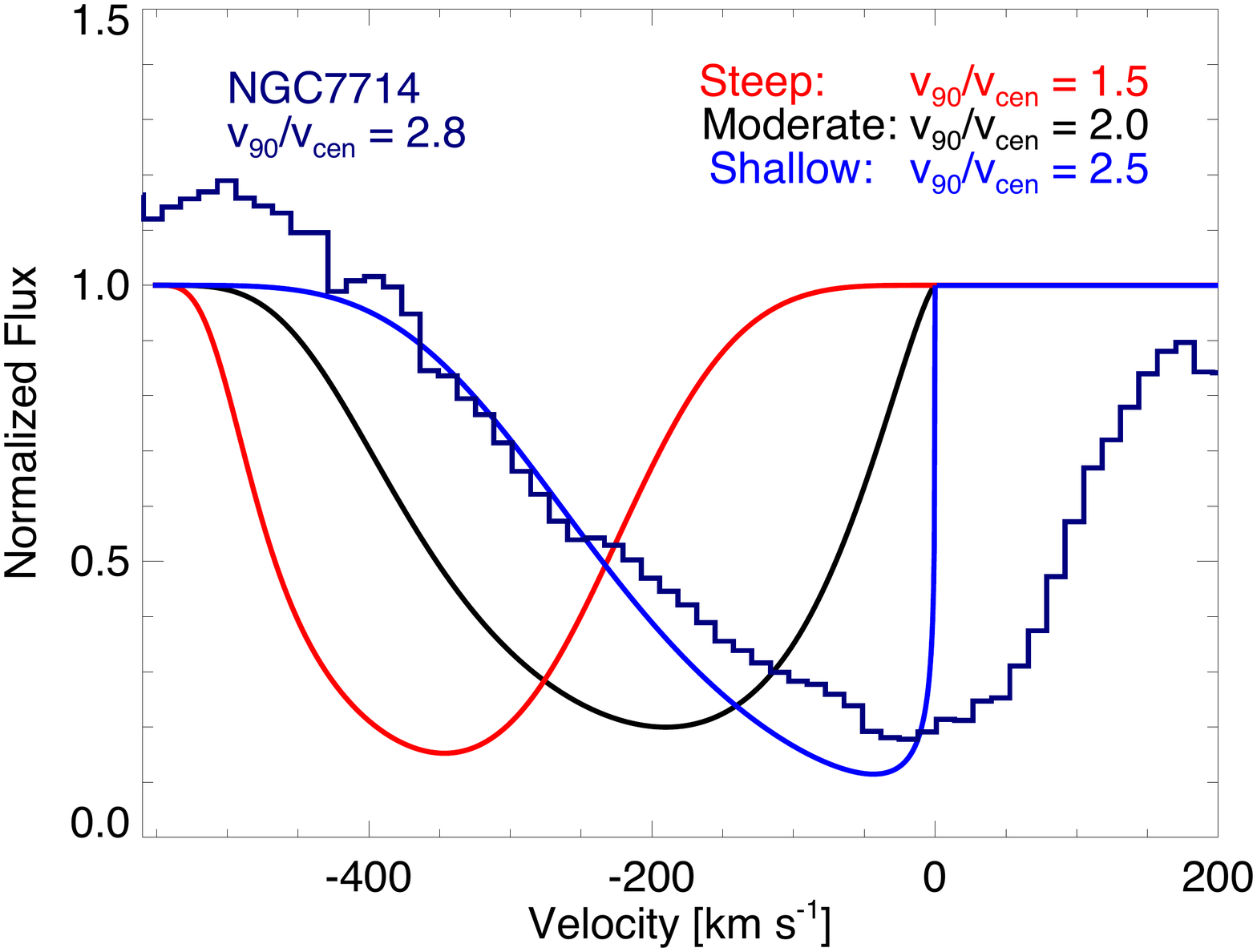}
\end{subfigure}
\begin{subfigure}{\textwidth}
\includegraphics[width = 0.5\textwidth]{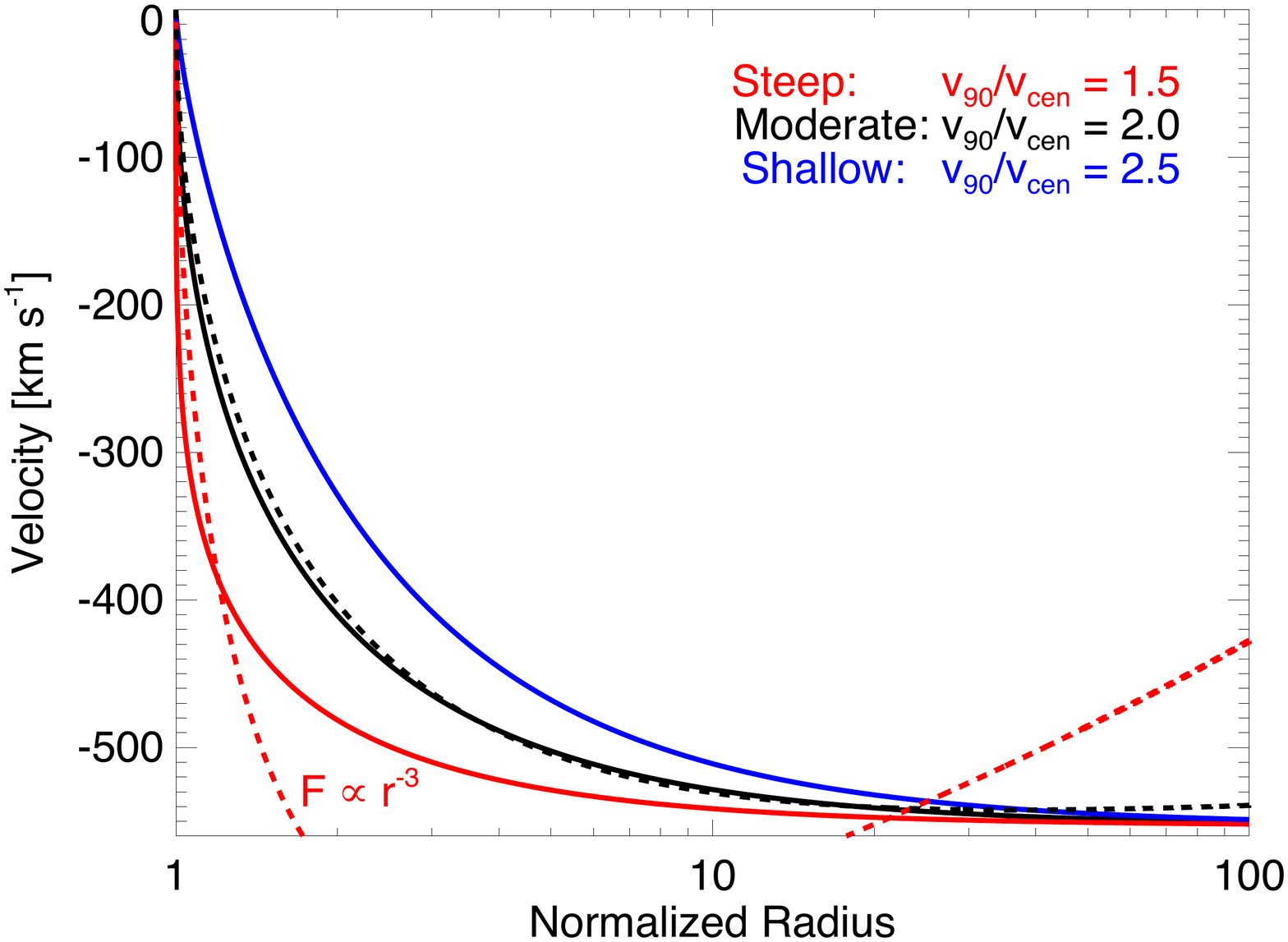}
\end{subfigure}
\caption{{\it Top Panel}: Theoretical line profiles for different acceleration profiles. The solid black curve corresponds to the observed profile from NGC~6090 \citep[see Fig.~\ref{fig:siprofiles};][]{Chisholm16b} while the red and blue curves correspond to steeply and shallowly accelerated outflows, respectively. These synthetic profiles correspond to a $\beta$ of 0.2, 0.43, and 0.75 for the steep, moderate, and shallow curves, respectively. If the outflow is rapidly accelerated, \vcen shifts closer to \vnp, as is observed in the leaker sample. Further, the measured outflow properties change with the acceleration profile: the \vn for the steep, moderate and shallow profiles are -367, -320 and -265~\kmsp, respectively, while the equivalent widths are 0.73, 0.84, and 0.74\AA. The dark-blue histogram data is the \siiv 1402\AA\ line profile from NGC~7714, which has a fitted $\beta = 0.84$ and a \vnp/\vcen of 2.8, consistent with the shallowly accelerated outflow. {\it Bottom panel}: The velocity versus normalized radius for the steep (red), moderate (black) and shallow (blue) acceleration profiles. In dashed lines are analytical relations for accelerating forces with an r$^{-2}$ force law (black dashed line) and an r$^{-3}$ force law (red dashed line).}
\label{fig:rat_comp}
\end{figure}

A $\beta$-law is a typical radial velocity profile for stellar winds because it describes accelerating an object as a radially dependent force \citep{Lamers}. The $\beta$-law says that the velocity is given as
\begin{equation}
v = v_\infty \left(1-\frac{\mathrm{R}_\mathrm{i}}{\mathrm{r}}\right)^\beta
\label{eq:beta}
\end{equation}
where $v_\infty$ is the maximum velocity and R$_\mathrm{i}$ is the initial radius of the outflow. The $\beta$ parameter determines the rate of acceleration, with $\beta \sim 0.5$ corresponding to an r$^{-2}$ force law in opposition to gravity. \citet{Chisholm16b} find the \siiv line profile from NGC~6090 corresponds to a $\beta$-law with $\beta = 0.4$. This line profile is plotted as the black curve in the upper panel of Fig.~\ref{fig:rat_comp}. The $\beta$-law from NGC~6090 produces a \vnp/\vcen ratio of 2. 

Additionally, in Fig.~\ref{fig:rat_comp} we show two different synthetic line profiles where we only change the $\beta$ value in Eq.~\ref{eq:beta}. In blue we show a slowly accelerating outflow ($\beta = 0.75$) where it takes the outflow nearly twice the radial distance to reach $-400$~\kms\ than it does for the $\beta =0.4$ profile (see the lower panel of Fig.~\ref{fig:rat_comp}). The velocity gradient (dr/d$v$) places the deepest part of the outflow at lower velocities, and creates a large blue tail with a \vnp/\vcen ratio of 2.5. Over-plotted in a dark-blue histogram is the \siiv absorption profile from NGC~7714, a large extended local starburst \citep{fox14}, which has a \vnp/\vcen ratio of 2.8, after accounting for significant zero-velocity absorption \citep{Chisholm15}. 

Finally, the red line in Fig.~\ref{fig:rat_comp} shows a rapidly accelerating outflow ($\beta $~=~$ 0.2$).  At low velocities the velocity changes so rapidly that dr/d$v$ drives Eq.~\ref{eq:sobop} to zero, even though the density is highest at low velocities. Eventually dr/d$v$ increases which sharply increases the optical depth. This steep velocity law has a narrow profile that peaks at relatively high velocities. This profile produces a small \vnp/\vcen ratio of 1.5, consistent with the small \vnp/\vcen values found for many of the leakers. This suggests that the small \vnp/\vcen ratios observed in the leakers could be because the leakers' outflows are steeply accelerated.  

The measured outflow properties change with different velocity laws. The shallowly accelerated outflow has a \vn of -265~\kms\ and an equivalent width of 0.74~\AA, while the rapidly accelerated profile has a \vn of -367~\kms\ and an equivalent width of 0.73~\AA, even though their density profiles are the same. Therefore, decreasing $\beta$ decreases \vn at a fixed equivalent width. This is similar to what we find in the upper right panel of Fig~\ref{fig:outflow}, where the leakers have an elevated \vn at fixed equivalent width.  

Physically, what does a steeper velocity gradient mean? A first possible explanation is that the accelerating force of the outflow scales differently with radius. In the lower panel of Fig.~\ref{fig:rat_comp} we show the radial velocity profiles for two possible force laws: 1) a force law that varies as r$^{-2}$ (black dashed line) and 2) a force law that varies as r$^{-3}$ (red dashed line). Each of these force laws include deceleration due to gravity. At all velocities, the r$^{-2}$ law matches the $\beta = 0.4$ acceleration profile \citep{Chisholm16b}. The r$^{-3}$ force law matches the $\beta = 0.2$ rapidly accelerated profile for the first 400~\kmsp, but the profile diverges at higher velocities as gravity decelerates the outflow. At the high velocities where the r$^{-3}$ profile diverges from the observations, the line profile is not observationally defined because the optical depth or covering fraction is low.

A second way to produce low \vnp/\vcen ratios is through a size correction of the driving force. In stellar wind theory, line driven winds accelerate gas off the stellar surface with a $\beta \sim 0.5$ \citep{Castor75}, but a finite size correction due to the size of the stellar photosphere steepens the velocity gradient to $\beta=0.8$ \citep{friend86}. Analogously, an extended galaxy has a larger finite size correction, creating a shallower velocity gradient and a larger \vnp/\vcen ratio. Since the leaking galaxies are extremely compact, the impact of a size correction could be less pronounced in these galaxies than more extended galaxies, producing more rapidly accelerated outflows.  Correspondingly, Haro~11, J0921$+$4509, and Tol~1247$-$381 are all fairly extended and have \vnp/\vcen ratios above 2.5, while the more compact galaxies have smaller \vnp/\vcen ratios. 

Rapidly accelerated outflows naturally explain why small \vnp/\vcen ratios lead to larger \fesclyp. The rapid acceleration moves neutral hydrogen out of the \lya\ reference frame, effectively reducing the optical depth. However, rapid acceleration does not guarantee that the Ly$\alpha$ photons escape because other factors like column density, metallicity and geometry still play a role in scattering Ly$\alpha$ photons. Therefore, small values of \vnp/\vcen produce a wide range of \fesclyp, depending on secondary parameters (column density, geometry), as seen in the middle panels of Fig.~\ref{fig:fesc}.

While the velocity profile drives the escape of \lya\ photons, the density profile should impact LyC escape. In Sect.~\ref{ew} we discuss how our observations favor a density-bounded scenario for escape, where the \hi\ density approaches zero before the ionizing photons are absorbed, but what factors lead to the density-bounded regime? We hypothesize that the outflow density follows a radial power-law density distribution as 
\begin{equation}
n (r) = n_0 \left(\frac{r}{R_\mathrm{i}}\right)^\alpha
\label{eq:dendist}
\end{equation}
where $n_0$ is the initial total hydrogen outflow density and $\alpha$ is the exponent that determines how rapidly the density decreases with radius. A value of $\alpha =-2$ corresponds to a density conserving sphere, while more negative values correspond to cases where density is not conserved and gas is removed from the outflow. Phase transitions or galactic fountains have been suggested to remove gas from the outflow, creating steep density profiles \citep{leroy15, Chisholm16b}. Consequently, sharp declines in the density with radius create the low-density conditions required for LyC photons to leak out of the galaxy. Accordingly, this affects the measured column density because
\begin{equation}
N  = \int_{R_\mathrm{i}}^\infty n\left(r\right) dr
\end{equation}
Using the radial density distribution from Eq.~\ref{eq:dendist}, we find that, if $\alpha < -1$, the \hi\ column density is
\begin{equation}
N_\mathrm{\ion{H}{i}} = -\frac{n_0 \chi_\mathrm{\ion{H}{i}} R_\mathrm{i}}{\alpha + 1}
\label{eq:column}
\end{equation}
where we have included the \hi\ ionization fraction ($\chi_\mathrm{\ion{H}{i}}$) to emphasize its role. To produce the small \hi\ column densities near 10$^{17}$~cm$^{-2}$ required for LyC escape, the initial density, the \hi\ ionization fraction, and the initial radius must be small, or the density must rapidly decline with radius (large, negative $\alpha$). Many of these properties are consistent with the leakers studied here: low \siii equivalent widths that are likely driven by small \hi\ column densities and metallicities (Sect.~\ref{ew}) from extremely compact (small R$_\mathrm{i}$) galaxies. The unsaturated \siiv absorption lines can be used to determine the $\alpha$ values, and recent work finds values between $-3$ and $-9$, with low-mass galaxies generally having more negative values (Chisholm et al. in preparation). Changing $\alpha$ from -2 to -9 decreases \vnp/\vcen from 2.07 to 1.97. While this is not a dramatic effect on the \vnp/\vcen ratio, it reduces the observed column density by a factor of 8 (see Eq.~\ref{eq:column}). This dramatically affects \fescp, and aids in truncating the \hi\ density to form a density-bound region.

The small \vnp/\vcen ratios of the leakers are reproduced with changes to the velocity gradient (dr/d$v$; Eq.~\ref{eq:beta}) and radial density profile (Eq.~\ref{eq:dendist}). Steep density-laws may create the density-bounded regions necessary to allow LyC photons to escape. The low \vnp/\vcen ratios also indicate that leakers have negligible zero-velocity absorption, which may enable the escape of ionizing photons from star-forming regions. The different line profiles underscore the importance of radially-resolved velocity and density profiles while discussing LyC and \lya\ escape. 

\subsection{Future steps and improvements}
\label{future}
The \vnp/\vcen ratio emphasizes the importance that the velocity and density profiles play for photons to escape galaxies. However, the \vnp/\vcen ratio is circumstantial evidence; direct evidence is required to determine their roles. The best direct evidence is to measure the velocity and density profile of unsaturated transitions, like \siivp~1402\AA, by simultaneously fitting the velocity-resolved optical depth and covering fraction distributions \citep{Chisholm16b}. Unfortunately, these measurements require higher signal-to-noise ratios, higher resolution, and redder wavelength coverage than these data afford, but future observations of the \siiv absorption profile from LyC leaking galaxies may determine the role of the velocity and density profiles. Additionally, hydrodynamic and radiative transfer simulations of clumpy media can diagnose why the \lya\ peak separation scales with \siii equivalent width.  

Another important consideration is how well the \siii transitions trace the neutral gas. In \citet{reddy} the \hi\ covering fraction is compared to the  \siiip~1260\AA\ covering fraction. Intriguingly, the authors find that the \siii absorption lines have covering fractions that are 2-3 times lower than the \hi. The authors suggest that the \siii absorption profiles might not probe \hi, implying that conclusions drawn from metal lines may not correspond to \hi\ gas. This effect deserves further examination to determine whether \siii directly relates to the \hi\ that absorbs LyC photons.  At high-resolution, this comparison can be made with \siii and \hi\ lines that are not contaminated by emission, like Ly$\beta$, to determine if they have similar covering fractions.  

Finally, the galaxies in the control sample do not have measured LyC escape fractions. To describe how the outflow velocities and equivalent widths influence the escape of ionizing photons, the community needs a statistical sample of galaxies that leak ionizing photons. A statistical sample would control for \fesc variations with outflow properties, rather than comparing leakers to potentially non-leaking galaxies. Current HST capabilities efficiently observe the LyC from galaxies at $z\sim0.2-0.3$, but COS has a finite lifetime, without a planned replacement in the next decade. Definitively answering how ionizing photons escape galaxies requires dedicated large surveys of leakers, while the facilities exist.

\section{Conclusions}
\label{summary}
Here we use a sample of nine local Lyman continuum (LyC) leaking galaxies with ultraviolet spectroscopy to test whether LyC leaking galaxies have extreme outflow velocities or strengths. We test this using a control sample of 27 local star-forming galaxies drawn from the {\sl Hubble Space Telescope} archive. We trace the galactic outflows using the equivalent widths and velocities of the \siiip~1260\AA\ and \siiiip~1206\AA\ absorption features, which probe neutral and ionized gas, respectively. In order to increase the dynamic range of the \hi\ column density probed, we also include a sample of double peaked \lya\ emitters. 

We find that galactic outflows from leakers have similar outflow velocities and smaller equivalent widths than the control sample (Fig~\ref{fig:outflow}). These velocities and equivalent widths also follow the scaling relations set by the stellar mass and metallicities of their host galaxies (Fig.~\ref{fig:host}). These trends indicate that leakers do not have extreme outflows, relative to the control sample. 

We find that the leakers have statistically weaker \siii equivalent widths than the control sample (upper left panel of Fig.~\ref{fig:outflow}) and that the equivalent widths are correlated with the gas phase metallicity (bottom right panel of Fig.~\ref{fig:host}). Further, the \siii equivalent width is strongly correlated with the velocity separation of the \lya\ peaks (Fig.~\ref{fig:fesc}), which correlates with the \hi\ column density. All of this indicates that LyC escape from these galaxies is likely because the \hi\ gas truncates before the ionizing photons are absorbed, consistent with a density-bound scenario (Sect.~\ref{ew}). 

Additionally, the leakers' \siii central velocities are closer to their maximum velocities than the control sample's (see top right panel of Fig.~\ref{fig:host}). These different asymmetries indicate that the leakers have negligible zero-velocity absorption that  may be caused by different acceleration profiles, namely different force laws or a finite size correction  (Fig.~\ref{fig:rat_comp} and Sect.~\ref{rat}). This emphasizes the role that the velocity and density profiles play in the escape of photons from galaxies.

While the silicon absorption lines do not indicate that LyC leaking galaxies drive extreme outflows, they do provide important clues for how ionizing photons escape galaxies. Higher resolution data will provide the opportunity to disentangle the effects of covering fraction and to measure the acceleration profiles. Larger sample sizes are also required to robustly determine how ionizing photons escape galaxies in order to reionize the universe.

\begin{acknowledgements}

We thank the anonymous referee for careful and thoughtful suggestions that greatly improved the manuscript. We thank Alaina Henry for thoroughly reading the manuscript and providing helpful corrections. We are grateful that Nicha Leethochawalit and Tucker Jones shared their DEIMOS spectra of high redshift galaxies with us.
I.O. acknowledges support of the Czech Science Agency grant 17-06217Y. TXT acknowledges the support of NASA through grant number HST-GO-13744.001-A from the Space Telescope Science Institute, which is operated by AURA, Inc., under NASA contract NAS5-26555. A.V. is supported by a Marie Heim Vögtlin fellowship of the Swiss National Foundation. 

\\
This research made use of the NASA's Astrophysics Data System (ADS) Bibliographic Services, NASA/IPAC Extragalactic Database (NED), which is operated by the Jet Propulsion Laboratory, California Institute of Technology, under contract with the National Aeronautics and Space Administration, and the SDSS database. Funding for SDSS-III has been provided by the Alfred P. Sloan Foundation, the Participating Institutions, the National Science Foundation, and the U.S. Department of Energy Office of Science. The SDSS-III web site is http://www.sdss3.org/. SDSS-III is managed by the Astrophysical Research Consortium for the Participating Institutions of the SDSS-III Collaboration including the University of Arizona, the Brazilian Participation Group, Brookhaven National Laboratory, Carnegie Mellon University, University of Florida, the French Participation Group, the German Participation Group, Harvard University, the Instituto de Astrofisica de Canarias, the Michigan State/Notre Dame/JINA Participation Group, Johns Hopkins University, Lawrence Berkeley National Laboratory, Max Planck Institute for Astrophysics, Max Planck Institute for Extraterrestrial Physics, New Mexico State University, New York University, Ohio State University, Pennsylvania State University, University of Portsmouth, Princeton University, the Spanish Participation Group, University of Tokyo, University of Utah, Vanderbilt University, University of Virginia, University of Washington, and Yale University.

\end{acknowledgements}

\appendix
\section{Remeasuring \fesc}
\label{remfesc}
\begin{table*}
\caption{Remeasured \fescp}
\begin{tabular}{cccccccc}
\hline
\hline
Galaxy Name	&	$F_\mathrm{lit}$	& Range &	$F_\mathrm{total}$		&	$F_\mathrm{shadow}$		&	Range	& $f_\mathrm{esc,lit}^{\rm LyC}$& $f_\mathrm{esc,new}^{\rm LyC}$ \\
 & & (\AA) & & & (\AA) & \% & \% \\ \hline
(1)&(2)&(3)&(4)&(5)&(6)&(7)&(8) \\
\hline
J0921$+$4509 &	0.30$\pm$0.07 &	902-912 &	0.26$^{+0.03+0.00}_{-0.01-0.01}$ &	0.26$^{+0.02+0.00}_{-0.02-0.00}$ &	880-912 & $\sim1$ &		$\sim1$ \\
Tol0440$-$381 &	10.30$\pm$2.30 &	882-912 &	3.49$^{+0.64+0.03}_{-0.65-0.02}$ &	2.71$^{+1.66+0.00}_{-1.27-0.12}$ &	880-912 & <7.1	&	1.9$^{+1.1}_{-1.0}$ \\
Tol1247$-$232	& 22.60$\pm$6.20 &	882-912 &	2.80$^{+0.59+0.00}_{-0.54-0.03}$ &	2.04$^{+0.96+0.00}_{-0.76-0.08}$ &	880-912 & 4.5$\pm$1.2 &	0.40$^{+0.19}_{-0.17}$ \\
Mrk~54	&	9.76$\pm$2.94 &	882-912 &	1.44$^{+0.60+0.07}_{-0.70-0.00}$ &	<0.67  &880-912 & 2.5$\pm$0.72 &	<0.16 \\
\hline
\end{tabular}
\tablefoot{Column (1) gives the name of the galaxy, column (2) gives the measured LyC average flux level calculated in the range given in column (3). Column (4) gives the total LyC flux in our remeasured method, while column (5) gives the total flux in just the shadow.  Larger $F_\mathrm{total}$ than $F_\mathrm{shadow}$ indicates significant geocoronal Lyman series emission or scattered geocoronal Lyman alpha emission. Both column (4) and column (5) are calculated in the range given in column (6). Columns (7) and (8) are the \fesc fractions measured from $F_\mathrm{lit}$ and $F_\mathrm{shadow}$, respectively. The errors for $F_\mathrm{total}$ and $F_\mathrm{shadow}$ are given as 1$\sigma$ errors where we have broken out the statistical and systematic errors as the first and second terms, respectively. The fluxes are in units 10$^{-16}$~erg~cm$^{-2}$~s$^{-1}$~\AA$^{-1}$. They are not corrected for Galactic extinction. Note that MRK~54 is not considered a LyC leaker, and it is included in the control sample.}
\label{tab:fesc}
\end{table*}

\begin{figure}
\includegraphics[width = 0.5\textwidth]{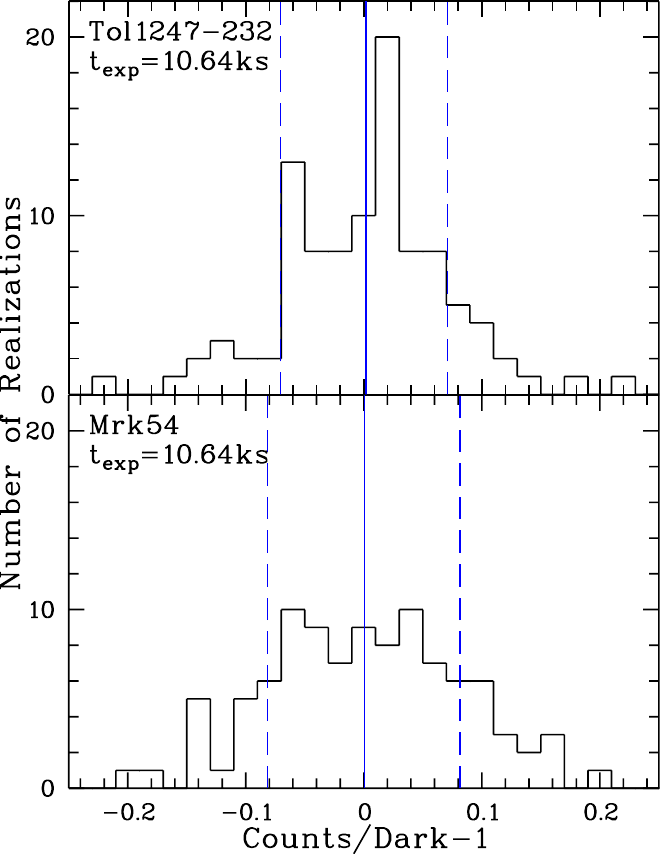}
\caption{Relative deviation for two galaxies (Tol~1247-232 and Mrk~54) between measured dark counts and estimated dark current for random samples of 8 of 65 dark exposures obtained in two 3-month intervals, using the same aperture and wavelength range as the science data (880-912\AA\ in the coadded data, including detector offsets). The solid and dashed lines show the mean and the standard deviation of the 100 realizations, respectively.}
\label{fig:append}
\end{figure}

We used our custom reductions of the archival leakers to reassess their LyC fluxes. We measured the mean fluxes in the rest frame wavelength range 880-912\AA\ using a maximum likelihood method, computed the statistical Poisson error, and estimated the systematic error from the propagated background subtraction error \citep{Worseck16, Izotov16b}. Table~\ref{tab:fesc} compares our measured fluxes to the values reported in \citet{Borthakur14} and \citet{Leitherer16}. For J0921$+4509$ we confirm the flux measurement from \citet{Borthakur14} with high confidence (10.2$\sigma$, calculated from Poisson distribution). The precision of the resulting \fescp$=1$\% is limited by assumptions in the SED modeling. For the three galaxies studied in \citet{Leitherer16} we measure significantly lower fluxes (a factor of 3--8). Although \citet{Leitherer16} use a technique similar to ours, they may have underestimated the dark current (their Figure 20). In fact, \citet{Puschnig} reanalyze the Tol~1247-232 data and find similarly decreased \fesc of $1.5\pm0.5$\%, after discovering that the {\small CALCOS 2.20d} pipeline fails at low/negative flux levels when superdark frames are included.   We show in Fig~\ref{fig:append} that our procedure estimates the COS dark current with negligible systematic errors. Moreover, we measure somewhat lower fluxes in shadows, indicating the presence of geocoronal Lyman series emission and/or scattered geocoronal Lyman alpha emission, although the statistical errors are large due to the smaller exposure times. Due to possible residual contamination from geocoronal Lyman series and scattered light, we regard the apparent detections of Tol~0440-381 and Tol~1247-232 in only the shadow data as upper limits. The shadow flux of Mrk54 is consistent with zero, and we do not consider these observations as a LyC detection. This emphasizes the difficulty of measuring LyC flux from very low-redshift galaxies because scattered geocoronoal Lyman series emission is easily confused for LyC emission. 

As in \citet{Worseck16}, we validated our dark estimation procedure by treating a subset of dark calibration exposures as data and estimating the dark current from the remaining dark calibration exposures. The COS dark monitoring program collects $5\times1330$~s of darks per week, or 65 exposures over a three-month interval, that we consider for the dark current estimation. Random subsamples of 8 exposures were drawn and reduced in the same fashion as the real data, accounting for detector offsets, while the remaining 57 exposures were taken for dark calibration. Total counts and estimated dark counts were measured on the same pixels as for the science data. However, the dark calibrations were infrequently observed, therefore it was impossible to match the science exposure times, or the environmental conditions, during the observations. Fig.~\ref{fig:append} shows the distributions of the relative deviations between the measured dark counts and the estimated dark current for sets of dark exposures taken within 3 months around the observation dates of two of the galaxies from \citet{Leitherer16}. From 100 realizations we find an insignificant systematic error of the estimated dark current and a relative error of $\sim8$\% that is dominated by the statistical Poisson error of the measured counts.

\bibliographystyle{aa} 
\bibliography{references} 

\end{document}